\documentclass[10pt,a4paper]{article}

\usepackage{amsmath,amssymb,amsthm,graphicx}
\usepackage{stfloats}
\usepackage{subfigure}

\usepackage{amsfonts, amsbsy, amssymb, amsmath, graphicx, float, subfigure, color}
\usepackage{hyperref}

\begin{document}

\author{J. Curbelo, A.M. Mancho}

\date {}

\title{Spectral numerical schemes for time-dependent  convection  with  viscosity dependent on temperature.}

\maketitle

\begin{abstract}
This article proposes  spectral  numerical methods to solve the time evolution of { convection problems  with  viscosity  strongly depending on temperature at infinite Prandtl number. Although we verify the proposed techniques just for viscosities that depend exponentially on temperature, the methods  are extensible to other dependence laws.}  The set-up is a 2D domain with periodic boundary conditions along the horizontal coordinate. { This introduces a symmetry in the problem, the O(2) symmetry,  which
is particularly well described by spectral methods and motivates the use of these methods in this context. We examine the scope of our techniques by  exploring   transitions from stationary regimes   towards time dependent regimes. At a given aspect ratio}
stable stationary solutions become unstable through a Hopf bifurcation, after which the time-dependent regime is solved by the  spectral techniques proposed in this article.

\end{abstract}




\section{Introduction}
\label{S1}
 {Thermal convection in fluids in which viscosity depends on temperature plays an important role in many geophysical and technical processes.
This problem 
has been addressed in the literature by considering diverse laws. For instance  an Arrhenius-type  viscosity law  is a usual approach to describe upper mantle convection problems \cite{KK97,ZP04,BMM92,Dav01}. Other studies such as \cite{TT71,SOB82, M82, BMM92, MS95} have considered fluids in which viscosity depends exponentially on temperature. In \cite{SOB82} an exponential law is chosen to fit the experimental data for the temperature dependence of viscosity in glycerol.
In \cite{BMM92} it is discussed the  exponential dependence as an approach to the Arrhenius law by means of a Taylor series  around a reference temperature, also called the Frank-Kamenetskii approximation (see \cite{FB02}). In \cite{MS95} extremely large viscosity variations as those expected in the mantle are investigated by means of an exponential law. More recently \cite{ULCRT12,CM13} have considered the hyperbolic tangent or the arctangent
as viscosity laws for they model a viscosity transition  in a narrow temperature gap. Other  studies have treated other weaker dependencies such as linear \cite{PEG67,RS93} or quadratic ones \cite{DS04,VW06}.}

 From the mathematical point of view   it has recently been proven  { that convection problems in which viscosity is a function of temperature} is a well posed problem \cite{GSW09,WZ11} for dependences which are smooth bounded positive analytical functions,  so it stands a good chance of solution on a computer using a stable algorithm. The variability in viscosity  introduces strong couplings
 between  the momentum and heat equations, as well as introducing important nonlinearities into the whole problem. On the other hand, it is of major interest, {in particular for mantle convection problems, }
 to consider  the fact that the Prandtl number, which is the quotient of viscosity and thermal diffusivity, is virtually infinite. This limit transforms the set of equations describing the time-dependent problem into a differential algebraic problem  (DAE), which is very stiff.

{In this context, this article  discusses  the performance of several time evolution spectral schemes for convection problems in which the viscosity  depends on temperature and the Prandtl number is infinite. We focus the analysis by choosing an exponential  law similar to that discussed in \cite{PMH09}. } We characterize  time-dependent solutions demonstrating the efficiency of the time-dependent scheme to describe the solutions beyond the stationary regime. {Our setting is a 2D domain with periodic boundary conditions along the horizontal coordinate. The equations with  periodic boundary conditions are invariant under horizontal translations, thus the problem has a symmetry represented by the  SO(2) group. Additionally,  if  the reflection symmetry exists, 
the full symmetry group is the O(2) group.
Symmetric systems typically exhibit more complicated behavior than non-symmetric systems and 
there exist numerous novel dynamical phenomena whose existence is fundamentally related to the presence of symmetry, such as traveling waves  or stable heteroclinic cycles \cite{ GH88,AGH88,PoMa97}. The numerical simulation of  dynamics under the presence of  symmetry has been usually addressed by spectral techniques \cite{PoMa97,HLB96}.
In the context of convection problems with constant viscosity in cylindrical containers that posses the O(2) symmetry,  the existence of
heteroclinic cycles have been reported both experimentally \cite{JN96} and numerically with a fully spectral approach \cite{DCJ00}. However   Assemat and co-authors \cite{ABK07} who have used high order finite element methods
to solve a similar set-up notice the absence of the heteroclinic cycles in their simulations. Additionally they notice the influence of the computational grid on the breaking of the O(2) symmetry, by producing pinning effects on the solution.
 These reasons suggest that spectral methods might  be  particularly suitable to deal with problems with  symmetries  as some solutions  might be overlooked with  approaches based on other spatial discretizations.}


Spectral methods {are not  very popular in the simulation of convection problems with temperature-dependent viscosity \cite{IT10}}, as they are reported to have limitations when handling lateral variations in viscosity.  Alternatively,  preferred schemes exist in which the basis functions are local;
for example, finite difference, finite element and finite volume methods. For instance, the works by \cite{G77,BI89}
have treated this problem  in a finite element discretization in primitive variables, while in \cite{C85,MPW74}  finite differences or finite elements  are used in the stream-function vorticity approach.
Spectral methods  have been  successfully applied to model mantle
convection with moderate viscosity variations in, 
for instance, \cite{CH91,CF03}. These works do not use the primitive variables formulation, and deal with the variations in  viscosity by  decomposing it  into a mean (horizontally averaged) part and
a fluctuating (laterally varying) part.  Our approach  addresses the variable viscosity problem by proposing a spectral approximation  in primitive variables without any decomposition on the viscosity.
 The main novelty in this paper is the extension of the spectral methodology discussed in    \cite{PMH09},  valid only for stationary problems,
 for solving the time-dependent problem, and the extension of the results  to describe   time-dependent solutions.

As regards temporal discretization,  backward differentiation formulas (BDF's) are widely used in convection problems.  This is the case
 of the work discussed in   \cite{MBA10}, which following ideas proposed in \cite{HuRa98} uses a  fixed time step second-order-accurate which combines   Adams-Bashforth and BDF schemes.
 A recent article  by Garc\'ia   and co-authors \cite{GNG-AS10} compares the performance  of several semi-implicit and implicit
 time integrations methods based on BDF and extrapolation formulas.
 The physical set-ups discussed in these papers are for convection problems with constant viscosity  and finite Prandtl number. In contrast, this article focuses on convection problems  with
  viscosity strongly dependent on temperature and infinite Prandtl number  that  lead to a differential algebraic problem. We will see that the semi-implicit methods discussed in  \cite{HuRa98,MBA10} do not work in this context.
  BDFs   and implicit methods are known to be an appropriate  choice \cite{HNW09, HaWa91} for efficiently tackling very stiff problems.  According to  \cite{HNW09, GNG-AS10},  for the time discretization scheme   we propose several high order
 backward differentiation formulas  which are ready for an automatic stepsize adjustment.
Furthermore,  we solve  the fully implicit problem and also propose a semi-implicit approach. The output and performance of this
  option are compared with those of the implicit scheme. It is found that  the semi-implicit approach presents some advantages in terms of computational performance.




The article is organized as follows: In Section \ref{S2}, we formulate the problem, providing a description of the physical set-up, the basic equations and the boundary conditions. Section \ref{SSolutions} describes a spectral scheme  for stationary  solutions
which will be  useful for  benchmarking the time dependent numerical schemes.  First the conductive solution and its stability is determined.
Other stationary solutions appear above the instability threshold, which are computed by means of a Newton-Raphson method using   a collocation method. The stability of  the stationary solutions is predicted by means of a linear stability analysis and is also solved with the spectral technique. 
Section   \ref{S3}  discusses several time-dependent schemes, which include  implicit and semi-implicit schemes.
  Section \ref{S4} reports the results at a fixed aspect ratio in a range of Rayleigh numbers. Stationary and time-dependent solutions are found and different morphologies of the thermal plumes are described. Some computational advantages of some schemes versus others are discussed. Finally, Section \ref{S5} presents the conclusions.

 \begin{figure}[!h]
 \centering
   \includegraphics[width=8.5cm]{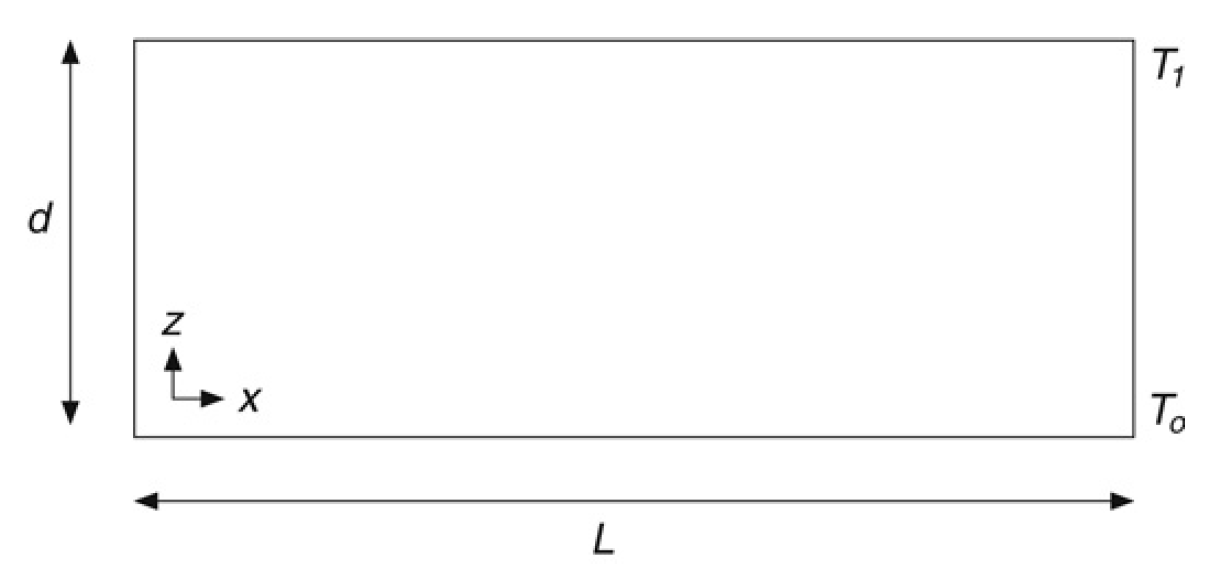}
    \caption{Problem set-up.}\label{F1}
   \end{figure}

 \section{Formulation of the problem}\label{S2}
   The physical set-up,  shown in Fig.\ref{F1},  consists of  a two dimensional fluid layer of depth $d$ ($z$ coordinate)  placed between two parallel plates of length $L.$   The bottom plate is at temperature $T_0$  and  the upper plate  is at $T_1,$ where $T_1=T_0-\Delta T$ and $\Delta T$ is the vertical temperature difference, which is positive, {\it i.e},  $T_1<T_0$.

   In the equations governing the system, $\mathbf{u}=(u_x,u_z)$ is the velocity  field, $T$ is the temperature, $P$ is the pressure, $x$ and $z$ are the spatial coordinates and $t$ is the time. Equations are simplified by taking into account the Boussinesq approximation, where the density $\rho$ is considered constant everywhere except in the external forcing term, where a dependence on temperature is assumed as follows $\rho=\rho_0(1-\alpha(T-T_1)).$ Here $\rho_0$ is the mean density at temperature $T_1$ and $\alpha$ the thermal expansion coefficient.

 We express the equations  with magnitudes in dimensionless form after
rescaling as  follows: $(x', z') = (x,z)/ d$, $t' = \kappa t / d^2$,
$\mathbf{u}'=  d \mathbf{u}/\kappa$, $P' = d^2 P / (\rho_0 \kappa \nu_0)$ , $\theta'= (T - T_1) /(\Delta T)$.
Here $\kappa$ is the thermal diffusivity and $\nu_0$ is the maximum viscosity of the fluid, which is the viscosity at temperature $T_1$.
After rescaling the domain, $\Omega_1 = [0,L)\times[0,d]$
is transformed into $\Omega_2 = [0 ,\Gamma)\times[  0 , 1], $ where
$\Gamma  = L / d$ is the aspect ratio.   The system evolves according to the momentum and the
mass balance equations, as well as to the energy conservation principle.  The non-dimensional equations  are (after dropping the primes in the fields):
   \begin{align}
&\nabla \cdot \mathbf{u}=0, \label{eqproblem1}\\
&\frac{1}{Pr}(\partial_t \mathbf{u}+ \mathbf{u}\cdot \nabla \mathbf{u}) = R\theta\vec{e}_3-\nabla P+\text{div} \left(\frac{\nu(\theta)}{\nu_0}(\nabla \mathbf{u}+(\nabla \mathbf{u})^T)\right),\label{eqproblem2}\\
&\partial_t \theta + \mathbf{u} \cdot \nabla \theta= \Delta \theta.\label{eqproblem3}
\end{align}
Here $\vec{e}_3$ represents the unitary vector in the vertical direction, $R=d^3 \alpha g \Delta T/(\nu_0 \kappa)$ is the Rayleigh number, $g$ is the gravity acceleration and $Pr=\nu_0/\kappa$ is the Prandtl number.
{We consider that the $Pr$ is  infinite, as is the case in mantle convection problems, and thus the term on the left-hand side in \eqref{eqproblem2} can be made equal to zero. This transforms the problem into a differential algebraic equation (DAE), which is very stiff.}

The viscosity $\nu(\theta)$ is a smooth positive  bounded function of $\theta$. {In order to test the performance of the proposed schemes  $\nu(\theta)$ is chosen to be an exponential law following previous results by \cite{PMH09}. The dimensional form of this law is as follows:
\begin{align}\label{eqexpd}
\frac{\nu(T)}{\nu_0}= \exp(-\gamma  (T-T_1))
\end{align}
where $\gamma$ is an exponential rate and $\nu_0$ is the largest viscosity at the upper surface. The dimensionless expression is:
\begin{align}\label{eqexp}
\frac{\nu(\theta)}{\nu_0}= \exp(-\mu R \theta)
\end{align}
where $\mu=\gamma \nu_0 \kappa/(d^3 \alpha g) $.  The presence of the $R$ number in the exponent of the viscosity law is uncommon among the literature that considers this viscosity dependence. However
it formulates better   laboratory experiments in which the increment of the $R$ number is done by increasing the temperature at the bottom  surface. This procedure  ties the viscosity to change
 with the Rayleigh number. }
 Dependence on Eq. (\ref{eqexp}) reduces to that  of constant viscosity if
$\mu=0$, while  for temperature-dependent viscosity  we  consider $\mu=0.0862.$ The viscosity contrast in equation  (\ref{eqexp}) for Rayleigh numbers up to $R=120$ --as employed in this article-- is $3.1 \cdot10^{4}$.

{Additionally  for benchmark purposes in section 3.1 the exponential law proposed by \cite{TT71} has been considered:
\begin{align}\label{eqexpt}
\frac{\nu(\theta)}{\nu_0}= \exp[c(\frac{1}{2}-\theta)]
\end{align}
where $c=\ln (\nu_{max}/\nu_{min})$. In this law $\nu_0$ is the viscosity involved in the  definition of the dimensionless $R$ number but it is not longer the maximum viscosity but the viscosity at $\theta=1/2$.
}

For boundary conditions, we consider that  the bottom plate is rigid and that the upper surface is non deformable and free slip.
The dimensionless boundary conditions are expressed as,
\begin{align}\label{eqbc}
\theta=1, \ \mathbf{u}=\vec{0},\text{ on } z=0 \text{ and } \
\theta=\partial_z u_x=u_z= 0,\text{ on } z=1.
\end{align}
Lateral  boundary conditions are periodic. Jointly with equations (\ref{eqproblem1})-(\ref{eqproblem3}), these conditions are invariant under translations along the $x$-coordinate, which introduces the symmetry SO(2) into the problem. In convection problems with constant viscosity, the reflexion symmetry $x\to -x$ is also present   insofar as  the fields are conveniently transformed as follows $(\theta,u_x,u_z,p) \to(\theta,-u_x,u_z,p)$. In this  case,   the O(2)  group expresses the full problem symmetry.
  The new terms introduced by the temperature dependent viscosity, in the current set-up Eq. (\ref{eqproblem2})  maintain the reflexion symmetry, and the symmetry group is O(2).

 \section{Numerical schemes for stationary solutions}\label{SSolutions}

The numerical codes proposed in this article for the time-dependent problem requiere {\it a priori} known solutions for benchmark.
In the set-up under consideration,  both stationary and time-dependent solutions exist. Some of the stationary solutions have simple analytical expressions which are known {\it a priori}, but there are others which are not so simple and must be found numerically. Stationary solutions are not stable in the full parameter space, and after becoming unstable either other stationary solutions may become stable or a time-dependent regime is observed.
In order to verify our methods, the self-consistency of results provided by a different kind of analysis is required. In this context, this section describes stationary solutions to the system (\ref{eqproblem1})-(\ref{eqproblem3})  and  their stability. This description, together with the fact  that the problem is well-posed \cite{WZ11}, provide a required  {\it a priori} knowledge that will assist in the examination of the validity of  different time-dependent numerical schemes.

\subsection{The conductive solution}
 The simplest stationary solution to the problem described by equations (\ref{eqproblem1})-(\ref{eqproblem3}) with boundary conditions (\ref{eqbc}) is the
  conductive solution which satisfies $\mathbf{u}_c=0$ and $\theta_c=-z+1$.  This solution is stable only  for a range of  vertical temperature gradients which are represented by
small enough Rayleigh numbers. Beyond the
critical threshold $R_c$,  a convective motion settles in and new structures are observed which
may be either time dependent or stationary.
The stationary equations in the latter case, obtained by canceling the time derivatives in the system  (\ref{eqproblem1})-(\ref{eqproblem3}), are satisfied
by the bifurcating solutions. As in the conductive solution the new solutions depend on the external
physical parameters, and new critical thresholds exist at which they lose their stability, thereby giving rise to new bifurcated structures.
     \begin{figure}[!h]
     \centering
   \includegraphics[width=8.5cm]{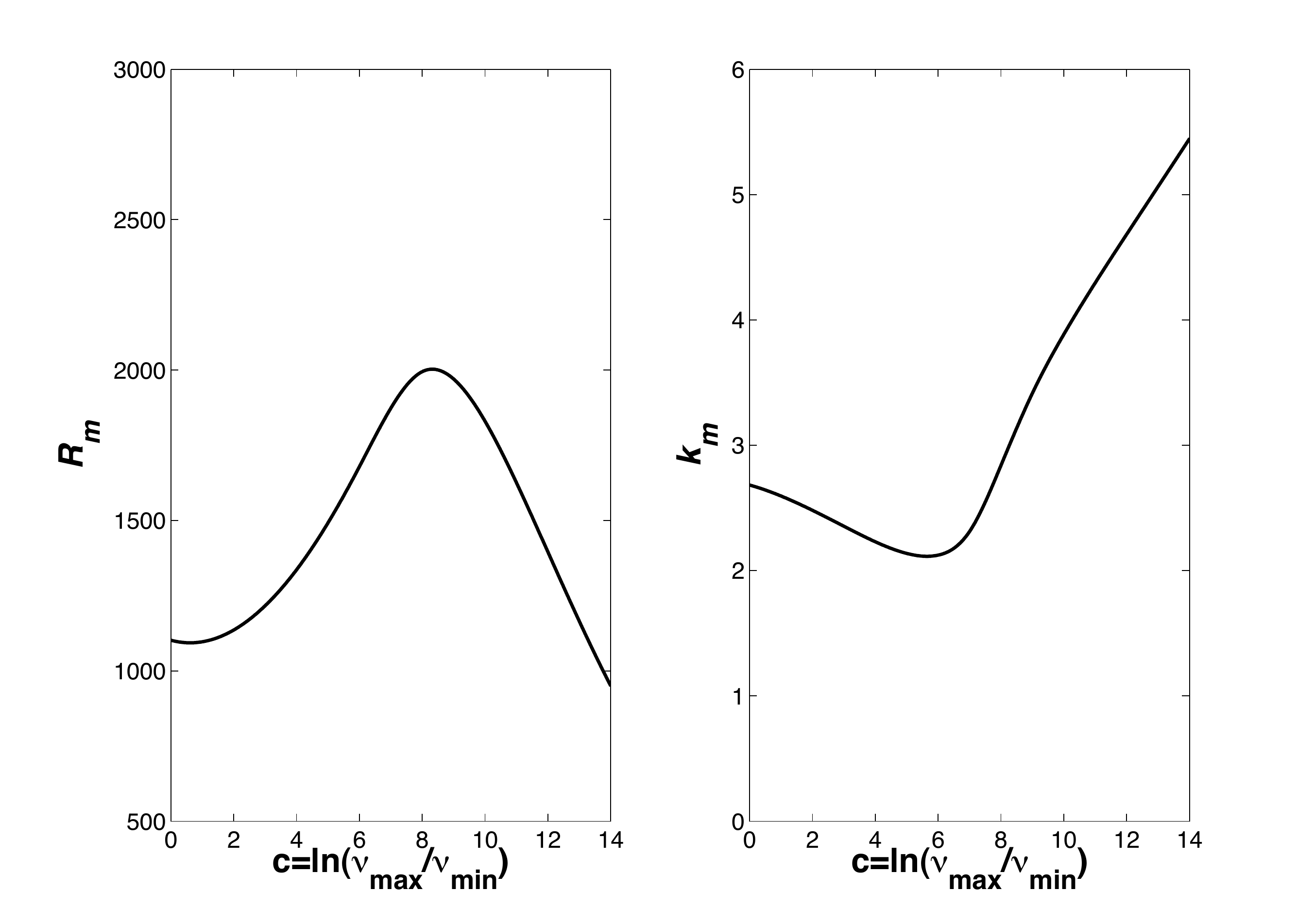}
    \caption{Critical curves according to the exponential law given in \eqref{eqexpt}. a) $R_m$ versus $c$; b) $k_m$ versus $c$.} \label{Fexpt}
   \end{figure}

     \begin{figure}[!h]
     \centering
   \includegraphics[width=8.5cm]{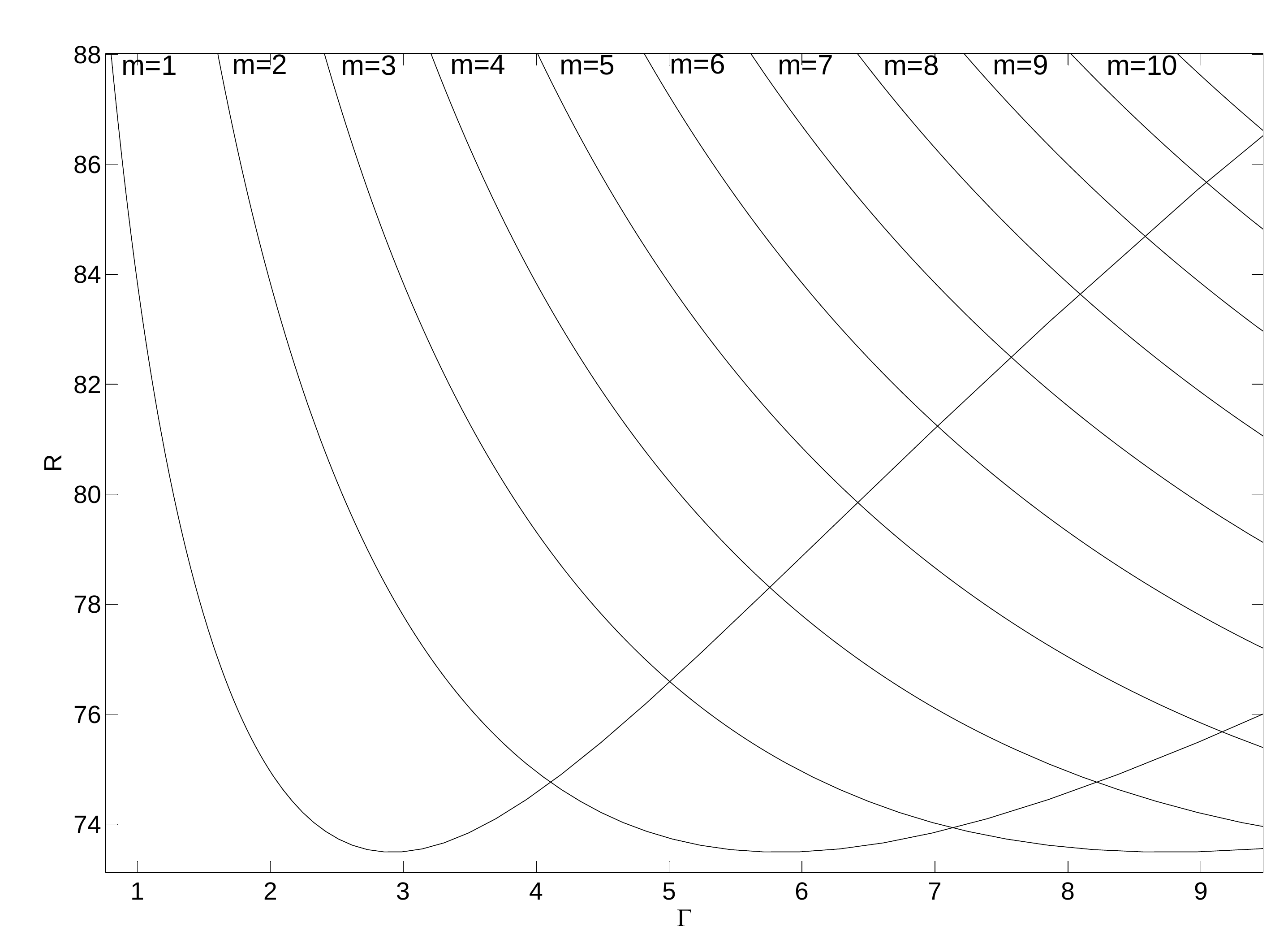}
    \caption{Critical instability curves $R(m,\Gamma)$ for a fluid layer with temperature dependent viscosity $\mu=0.0862$ according to the exponential law given in \eqref{eqexp}.} \label{Fexp}
   \end{figure}

In this section, we first describe the instability thresholds for the conductive solution. For this purpose small perturbations are added to it:
 \begin{eqnarray}
\mathbf {u}(x,z,t)&=&0+\tilde{\mathbf{u}} (z){\rm e}^{\lambda t+ikx},
\label{pert1}\\
\theta(x,z,t)&=&-z+1+\tilde{\theta}(z){\rm e}^{\lambda t+ikx},\label{pert2}\\
P(x,z,t)&=&-Rz^2/2+Rz+C+\tilde{P}(z){\rm e}^{\lambda t+ikx}.\label{pert3}
\end{eqnarray}
The sign in the real part of the eigenvalue $\lambda$ determines the stability of
the solution:  if it is  negative the perturbation decays and the stationary solution
is stable, while if it is positive the perturbation grows in time and the conductive solution
is unstable.
If these expressions are introduced into the system (\ref{eqproblem1})-(\ref{eqproblem3}), and both the nonlinear terms in the perturbations and their tildas are dropped,
the system becomes:
\begin{eqnarray}
0&=& iku_x+\partial_z u_z, \nonumber\\
0&=&ikP-\left[  \frac{\partial_z \nu(\theta_c)}{\nu_0} (iku_z+\partial_z u_x)  +     \frac{\nu(\theta_c)}{\nu_0} (\partial_{zz}^2 - k^2) u_x   \right], \nonumber\\
0&=&\partial_z P-\left[   2    \frac{\partial_z \nu(\theta_c)}{\nu_0}\partial_z u_z+       \frac{\nu(\theta_c)}{\nu_0} (\partial_{zz}^2 - k^2) u_z  \right] -R \theta,  \nonumber \\
\lambda \theta&= &(\partial_{zz}^2 -k^2 ) \theta + u_z. \nonumber
\end{eqnarray}
The boundary conditions for the perturbation fields are:
\begin{align}\label{eqbcp}
\theta=0, \ \mathbf{u}=\vec{0},\text{ on } z=0 \text{ and } \
\theta=\partial_z u_x=u_z= 0,\text{ on } z=1.
\end{align}
Stability analysis  of the conductive solution for  viscosity dependent on temperature
is numerically addressed   in \cite{SOB82,PMH09,PHL10}.  {For the results reported in this section we  follow the spectral scheme presented in \cite{HeMa02}. Appropriate expansions in Chebyshev polynomials of the unknown fields  $({\bf u}, \theta, P)$ along the vertical coordinate}  transform the eigenvalue problem into its discrete  form:
 $$ Aw=\lambda Bw,$$
where  the expansion coefficients are stored in the vector $w$. This generalized eigenvalue problem supplies  the  dispersion relation $R=R(k)$ at the bifurcation point (Re($\lambda)=0$).
{  $R(k)$ is  an upwards concave curve that reaches  a minimum at $R_m$, $k_m$. In an infinite domain.  $R_m$ and $k_m$  are referred respectively as the critical Rayleigh number and the critical wave number because above $R_m$ the conductive solution losses its stability and a new pattern
grows with wave number  $k_m$. For the viscosity law (\ref{eqexpt}) Stengel and co-authors \cite{SOB82}  have computed the values of $R_m$, $k_m$  as a function of the viscosity contrast $c=\ln(\nu_{max}/\nu_{min})$ ranging from
0 to 14. For  free slip conditions at the top and rigid conditions at the bottom,  their Figure 2  shows a curve that we reproduce with our spectral scheme also in our Figure \ref{Fexpt}. The agreement among them is excellent
and this provides a first benchmark for our   calculations.}

{In  finite domains   the wavenumber $k$  that appears} in the dispersion relation cannot be arbitrary but must meet the periodic boundary conditions, {\it i. e.}:
\begin{equation}
k \, \Gamma=2  \pi m, \,\,\, {\rm with} \,\,\, m=1,2,3\dots\label{wavenumber}
\end{equation}
Here, $m$ is  the number of wavelengths of the unstable structure
growing in the finite domain.
Restrictions to $k$ given by condition (\ref{wavenumber}) are replaced in the dispersion relation $R=R(k)$, thereby providing
a critical curve  for each integer $m$ as a function of the aspect ratio $\Gamma$, $R=R(m,\Gamma)$.
Figure  \ref{Fexp} displays  critical Rayleigh numbers $R_c$  on the vertical axis
as a function of the aspect ratio $\Gamma$ on the horizontal axis { for the viscosity law (\ref{eqexp}), which is the one we keep in the remaining analysis}. The conductive solution is stable below the critical curves, which means that in a box with a given aspect ratio, $\Gamma$,  if $R<R_c$
then initial conditions  near to  the conductive solution evolve in time  approaching it.
Alternatively, if at that aspect ratio $R>R_c$, then  initial conditions which are near to  the conductive solution
evolve in time  away from it,  towards a different  solution. This new solution may be stationary or time-dependent.
Figure \ref{Fexp} confirms that for increasing  aspect ratios the most unstable spatial eigenfunction increases its wavenumber $m$.

\subsection{Numerical stationary solutions}

 There exist  stationary solutions  to the system (\ref{eqproblem1})-(\ref{eqproblem3}) that bifurcate from the conductive solution above the instability thresholds
 displayed in Figure \ref{Fexp} and  may be computed numerically. They are stationary because they satisfy the stationary version of  equations (\ref{eqproblem1})-(\ref{eqproblem3}), which are obtained by canceling the partial derivatives with respect to time. These solutions may be numerically obtained  by using  a variant of the iterative Newton-Raphson method, similar to that in \cite{PMH09}. This method starts  with an  approximate solution at step $s=0$,  to which
 is added a small correction in tilda:
 \begin{equation}
 ({\bf u}^s+ \tilde{{\bf u}},  \theta^s + \tilde{\theta},  P^s + \tilde{P}). \label{pertu}
 \end{equation}
These expressions are introduced into the system (\ref{eqproblem1})-(\ref{eqproblem3}), and after canceling the nonlinear terms in tilda, the following equations are obtained:
\begin{align}
0=&\nabla \cdot \tilde{\mathbf{u}}+\nabla \cdot \mathbf{u}^s,\label{eqtem13}\\
0=&-\partial_x \tilde{P} -\partial_x P^s +\frac{1}{\nu_0} [L_{11}(\theta^s, u_x^s,u_z^s)+L_{12}(\theta^s)\tilde{u}_x \nonumber\\
&+L_{13}(\theta^s)\tilde{u}_z +L_{14}(\theta^{s},u_x^s, u_z^s)\tilde{\theta}],\label{eqtem23}\\
0=&-\partial_z \tilde{P}-\partial_z P^s+\frac{1}{\nu_0} [L_{21}(\theta^s, u_x^s,u_z^s)+L_{22}(\theta^s)\tilde{u}_x \nonumber\\
&+L_{23}(\theta^s)\tilde{u}_z +(L_{24}(\theta^{s},u_x^s, u_z^s)+R)\tilde{\theta}],\label{eqtem33}\\
0=&\tilde{\mathbf{u}}\cdot \nabla \theta^{s}+\mathbf{u}^{s}\cdot \nabla \tilde{\theta}+\mathbf{u}^{s}\cdot \nabla \theta^{s}-\Delta \tilde{\theta} -\Delta {\theta^s}.\label{eqtem43}
\end{align}
Here, $L_{ij}$ ($i=1,2$, $j=1,2,3,4$) are linear operators with non constant coefficients which are defined as follows:
\begin{align}
L_{11}(\theta, u_x, u_z)=&2 \partial_{\theta} \nu(\theta) \partial_x \theta \partial_x u_x + \nu(\theta)\Delta u_x\nonumber\\ &+ \partial_{\theta}\nu(\theta) \partial_z \theta (\partial_x u_z + \partial_z u_x),\label{eqL11}\\
L_{12}(\theta)=& 2\partial_{\theta} \nu(\theta)\partial_x \theta \partial_x + \nu(\theta)\Delta + \partial_{\theta} \nu(\theta)\partial_z \theta \partial_x ,\label{eqL12}\\
L_{13}(\theta)=& \partial_{\theta}\nu(\theta)\partial_z \theta\partial_x,\label{eqL13}\\
L_{14}(\theta,u_x,u_z)=& 2\partial_{\theta} \nu(\theta)\partial_x u_x\partial_x +2\partial^2_{\theta \theta}\nu(\theta) \partial_x \theta \partial_x u_x +\partial_{\theta}\nu(\theta) \Delta u_x \nonumber\\&+  (\partial_x u_z + \partial_z u_x)(\partial_{\theta} \nu(\theta)\partial_z +\partial^2_{\theta\theta}\nu(\theta)\partial_z \theta ),\label{eqL14}\\
L_{21}(\theta, u_x, u_z)=&2 \partial_{\theta} \nu(\theta) \partial_z \theta \partial_z u_z + \nu(\theta)\Delta u_z\nonumber\\ &+ \partial_{\theta}\nu(\theta) \partial_x \theta (\partial_z u_x + \partial_x u_z),\label{eqL21}\\
L_{22}(\theta)=&\partial_{\theta}\nu(\theta)\partial_x \theta\partial_z ,\label{eqL22}\\
L_{23}(\theta,u_x,u_z)= & 2\partial_{\theta} \nu(\theta)\partial_z \theta \partial_z + \nu(\theta)\Delta + \partial_{\theta} \nu(\theta)\partial_x \theta \partial_z ,\label{eqL23}\\
L_{24}(\theta,u_x,u_z)=& 2\partial_{\theta} \nu(\theta)\partial_z u_z\partial_z  +2\partial_{\theta \theta}\nu(\theta) \partial_z \theta \partial_z u_z +\partial_{\theta}\nu(\theta)\Delta  u_z  \nonumber\\
&+  (\partial_z u_x + \partial_x u_z)(\partial_{\theta} \nu(\theta)\partial_x +\partial_{\theta\theta}\nu(\theta)\partial_x \theta).\label{eqL24}
\end{align}
 {In the above expressions spatial derivatives of the viscosity function $\nu(\theta)$ are computed through the chain rule as numerically  this provides more accurate results}. The unknown fields
$\tilde{{\bf u}}, \,\, \tilde{P}, \,\, \tilde{\theta}$ are found by solving the linear system with the boundary conditions (\ref{eqbcp})
and the new approximate solution $s+1$ is  set to
\begin{equation} {\bf u}^{s+1}= {\bf u}^s+ \tilde{{\bf u}}, \;  \theta^{s+1}= \theta^s+ \tilde{\theta}, \;  P^{s+1}= P^s+ \tilde{P}.\nonumber\end{equation}
The whole procedure is repeated for $s+1$ until a convergence criterion  is fulfilled. In particular,
we  consider
that the $l^{2}$ norm of the computed perturbation
should be less than $10^{-9}$.


At each step, the resulting linear system  is solved by
expanding any unknown perturbation field $Y,$ in Chebyshev polynomials in the vertical direction and Fourier modes along the horizontal axis:
\begin{align}\label{eqexpansion}
Y(x,z)=&\sum_{l=1}^{\lceil L/2\rceil}\sum_{m=0}^{M-1} a^Y_{lm}T_m(z)e^{i(l-1)x} + \sum_{l=\lceil L/2\rceil+1}^{L}\sum_{m=0}^{M-1} a^Y_{lm}T_m(z)e^{i(l-1-L)x}.
\end{align}
In this  notation,  $\lceil \cdot \rceil$ represents the nearest integer towards infinity. Here $L$ and $M$ are the number of nodes in the horizontal and vertical directions, respectively.
Chebyshev polynomials are defined in the interval $[-1,1]$ and Fourier modes
in the interval $[0,2  \pi].$ Therefore, for computational convenience, the domain $\Omega=[0,\Gamma)\times[0,1]$ is transformed into $[0,2\pi)\times[-1,1].$
 This change in coordinates introduces
scaling factors into equations and boundary conditions which are not
explicitly given here.
There are $4\times L\times M$ unknown coefficients  which are determined by a collocation method in which equations (\ref{eqtem13})-(\ref{eqtem33}) and boundary conditions
are posed at the collocation points $(x_j,z_i),$
\begin{align*}
\text{Uniform grid:}\quad &x_j= (j-1)\frac{2\pi}{L}, &j=1,\ldots, L;&\\
\text{Gauss--Lobatto:} \quad &z_i= \cos\left( \left( \frac{i-1}{M-1}-1\right)\pi\right), & i=1,\ldots,M;&
\end{align*}
After replacing expression (\ref{eqexpansion})  in equations   (\ref{eqtem13})-(\ref{eqtem33}), the partial derivatives are evaluated on the basis of functions. Derivatives of the Chebyshev polynomials at the collocation points  are evaluated {\it a priori} according to the ideas reported in \cite{HM00}.
The expansion (\ref{eqexpansion}) is an  interpolator outside the collocation  points, which when restricted to these points may be rewritten as:
\begin{align}\label{eqexpansion2}
Y(x_j,z_i)=&\sum_{l=1}^{L}\sum_{m=0}^{M-1} a^Y_{lm}T_m(z_i)e^{i(l-1)x_j},
\end{align}
due to the aliasing effect  of   the functions $e^{i(l-1)x_j}$ and $e^{i(l-1-L)x_j}$ for $l>L/2$ at the collocation points. However, this expression is not valid  for computing the spatial derivatives of the fields,  for which purpose
expansion (\ref{eqexpansion}) should be used. Although in practice Eq. (\ref{eqexpansion})  is  correct and provides good results, we do not employ it in this work because it involves complex functions and complex unknowns  $a^Y_{lm}$, eventually   leading to the inversion of complex matrices which
computationally are more costly than real matrices. On the other hand since the unknown functions $Y$ are real, they admit expansions with real functions
and real unknowns. In order to obtain these functions, we take into account Euler's formula $e^{ilx}=\cos(lx)+i\sin(lx),$ which is replaced in \eqref{eqexpansion}. We also note
 that the coefficients are given by conjugated pairs in such a way that, for instance, $a_{2m}^{Y}=a_{Lm}^{Y*}$.  Strictly speaking, expansion (\ref{eqexpansion}) is a real function
 only if every coefficient in the first summatory for $l\ge 2$ has a conjugate pair in the second summatory. This implies that $L$ must be an odd number; thus in what follows we restrict ourselves to odd $L$ values. With these considerations the following equations are obtained:
\begin{align}\label{eqexpansion3}
Y(x,z)=&\sum_{l=1}^{\lceil L/2\rceil}\sum_{m=0}^{M-1} b^Y_{lm}T_m(z)\cos((l-1)x) \nonumber \\ &+ \sum_{l=2}^{\lceil L/2\rceil}\sum_{m=0}^{M-1} c^Y_{lm}T_m(z)\sin((l-1)x).
\end{align}
Some relations among the real and complex coefficients are: $b^Y_{1m}=a^Y_{1m}$ and $b^Y_{lm}=2 \Re(a^Y_{lm})$ and $c^Y_{lm}=-2 \Im(a^Y_{lm})$, for $l=2,\dots, \lceil L/2 \rceil.$

The rules followed to obtain as many equations as unknowns are described next. Equations  \eqref{eqtem13}--\eqref{eqtem43} are evaluated at nodes $i=2,\dots, M-1$, $j=1,\dots,L.$ This provides $4\times (M-2)\times L$ equations;  the boundary conditions \eqref{eqbc} are evaluated at $i=1,M$, $j=1,\dots,L.$ This supplies additional  $4\times L \times M-2L$ equations. In order to obtain the remaining $2L$ equations, we complete the system with extra boundary conditions which eliminate spurious modes for pressure \cite{HeMa02,HHDMCPY03} projecting the equation of motion in the upper and lower plate of the domain {\it  i.e.} the equation \eqref{eqtem33} is evaluated at nodes $i=1,M$, $j=1,\dots,L.$ This choice has been reported to be successful for many convection problems \cite{HeMa02, HHM02b, NMH07}. However, in the present set-up, results are improved if,  for the equation \eqref{eqtem33} imposed at the upper boundary,
 the continuity equation is assumed and  $\partial^2_{zz}u_z $ is replaced with $-\partial_{xz}u_x.$
 With these rules  we obtain a  linear system of the form $ AX = b $ in which $X$ contains the unknowns. However, the matrix $A$ is singular due to the fact that pressure with the imposed conditions is defined up to an additive constant. According to\cite{HeMa02, HHM02b, NMH07}, we fix the constant
by removing
 equation \eqref{eqtem33} at node $j=1$, $i=2$  and adding  at this point the equation $b_{10}^P=0$.  This is computationally  cheaper than the pseudo-inverse method. However,
 in the problem under study, dropping the   equation \eqref{eqtem33}
 at one point introduces weakly oscillating structures on this side of  the pressure field.
To overcome this drawback, in the final step of the iterative procedure, once the tolerance is attained,
we proceed alternatively by computing a pseudoinverse of the matrix $A$ by using the singular value decomposition (SVD). Let $A = U\Sigma V^*$ be the singular value decomposition of $A$; $V\Sigma^+U^*$ is the pseudoinverse of A, where $\Sigma^+$ is the pseudoinverse of a diagonal matrix, i.e. it takes the reciprocal of each non-zero element on the diagonal, and transposes the resulting matrix.


The study of the stability  of the numerical stationary solutions under consideration is  addressed  
 by means of a linear stability analysis. Now perturbations are added to a general
stationary solution, labeled with superindex $b$:
\begin{eqnarray}
\mathbf {u}(x,z,t)&=&\mathbf {u}^b(x,z)+\tilde{\mathbf{u}} (x,z){\rm e}^{\lambda t}
\label{pert1},\\
\theta(x,z,t)&=&\theta^b(x,z)+\tilde{\theta}(x,z){\rm e}^{\lambda t},\label{pert2}\\
P(x,z,t)&=&P^b(x,z)+\tilde{P}(x,z){\rm e}^{\lambda t}.\label{pert3}
\end{eqnarray}
The linearized equations are:
\begin{align}
0=&\nabla \cdot \tilde{\mathbf{u}} \label{eqtem13_2}\\
0=&-\partial_x \tilde{P}  +\frac{1}{\nu_0} [L_{12}(\theta^b)\tilde{u}_x +L_{13}(\theta^b)\tilde{u}_z +L_{14}(\theta^{b},u_x^b, u_z^b)\tilde{\theta}]\label{eqtem23_2}\\
0=&-\partial_z \tilde{P}+\frac{1}{\nu_0} [L_{22}(\theta^b)\tilde{u}_x+L_{23}(\theta^b)\tilde{u}_z +(L_{24}(\theta^{b},u_x^b, u_z^b)+R)\tilde{\theta}]\label{eqtem33_2}\\
0=&\tilde{\mathbf{u}}\cdot \nabla \theta^{b}+\mathbf{u}^{b}\cdot \nabla \tilde{\theta} +\mathbf{u}^{b}\cdot \nabla {\theta^b}-\Delta \tilde{\theta}+\lambda \tilde{\theta},\label{eqtem43_2}
\end{align}
where the operators $L_{ij}$ are the same as those defined in Eqs. (\ref{eqL11})-(\ref{eqL24}). The stability of the stationary solutions is approached with the  collocation method
used  for the Newton-Raphson iterative method. Expansions  of the fields (\ref{eqexpansion3}) are replaced in equations (\ref{eqtem13_2})-(\ref{eqtem43_2}), and they and the boundary conditions (\ref{eqbcp}) are evaluated at the  collocation nodes following the same rules as before. As a result,  the discrete form of the generalized eigenvalue problem is obtained:
\begin{align}\label{eig} Aw=\lambda Bw,\end{align}
 where $w$ is a vector containing unknowns.

In order to solve the generalized eigenvalue problem, we use a generalized Arnoldi method, which is described in  \cite{NHMW08}. The numerical approach uses the idea of preconditioning the eigenvalue problem with a modified Caley transformation, which transforms the problem \eqref{eig} and which admits infinite eigenvalues into another one with all its eigenvalues finite. Afterwards,  the Arnoldi method is applied.

 \section{Numerical schemes for time dependent solutions}
 \label{S3}

The governing equations \eqref{eqproblem1}--\eqref{eqproblem3}, together with boundary conditions \eqref{eqbc}, define a time-dependent problem for which we discuss temporal schemes based on  a primitive variables formulation.  The spatial discretization is  analogous to that proposed in the previous section, thus
the focus in this section is to discuss the time discretization of the problem.

To integrate in time, we use a  third order multistep scheme. In particular,  we use a backward differentiation formula (BDF), since as discussed in  \cite{HNW09,HaWa91}
 these are highly appropriate for very stiff problems such as ours. The BDF  evaluates the time derivative in Eq. \eqref{eqproblem3} by differentiating
 the formula that extrapolates the field $\theta^{n+1}$ with a third order Lagrange polynomial  that uses fields at times  $\theta^{n}, \theta^{n-1}, \theta^{n-2}$, {\it i.e.}:
 $$ \theta(t):=\ell_{n+1}(t)\theta^{n+1}+\ell_{n}(t)\theta^{n} +\ell_{n-1}(t)\theta^{n-1} +\ell_{n-2}(t)\theta^{n-2} $$
  $\ell_k$ being a Lagrange polinomial of order 3:
 $$
 \ell_{k}(t)=\prod_{  \begin{array}{c} n-2\leq m\leq n+1 \\ m \neq k  \end{array}}  \frac{t-t_m}{t_{k}-t_m}, \quad n-2 \leq k\leq n+1
 $$
 Then,
 \begin{equation}\label{eqdttheta} \partial_t \theta^{n+1}= \ell_{n+1}'(t_{n+1})\theta^{n+1} + \ell_{n}'(t_{n+1})\theta^n+ \ell_{n-1}'(t_{n+1})\theta^{n-1}+\ell_{n-2}'(t_{n+1})\theta^{n-2}    \end{equation}
For the fixed time step case, the time differentiation  simplifies to the equation:
\begin{equation}\label{eqbdffs}
\partial_t \theta^{n+1}= \frac{11\theta^{n+1}-18\theta^n+ 9\theta^{n-1}-2\theta^{n-2}}{6\Delta t}
\end{equation}
In stiff problems, a variable time step scheme with an adaptative step control that adjusts itself conveniently to the different regimes is advisable.  The expressions for the time derivative  \eqref{eqdttheta} for the variable time step  case are:
 \begin{align*}
\ell_{n+1}'(t_{n+1})=& \frac{\Delta t_n^2+4\Delta t_{n+1}\Delta t_n+\Delta t_{n}\Delta t_{n-1}+3\Delta t_{n+1}^2+2\Delta t_{n-1}\Delta t_{n+1}}{(\Delta t_{n}+\Delta t_{n+1}+\Delta t_{n-1})(\Delta t_{n}+\Delta t_{n+1})\Delta t_{n+1}}\\
\ell_{n}'(t_{n+1})=&\frac{\Delta t_{n}^2+\Delta t_{n}\Delta t_{n-1}+\Delta t_{n-1}\Delta t_{n+1}+\Delta t_{n+1}^2+2\Delta t_{n}\Delta t_{n+1}}{(\Delta t_{n}+\Delta t_{n-1})\Delta t_{n}\Delta t_{n+1}}\\
\ell_{n-1}'(t_{n+1})=&\frac{\Delta t_{n+1}(\Delta t_{n}+\Delta t_{n+1}+\Delta t_{n-1})}{(\Delta t_{n}+\Delta t_{n+1})\Delta t_{n}\Delta t_{n-1}} \\
\ell_{n-2}'(t_{n+1})=& \frac{\Delta t_{n+1}(\Delta t_{n}+\Delta t_{n+1})}{((\Delta t_{n}+\Delta t_{n-1})^2+\Delta t_{n}\Delta t_{n+1}+\Delta t_{n-1}\Delta t_{n+1})\Delta t_{n-1}},
\end{align*}
 where $\Delta t_{n}=t_{n}-t_{n-1}.$

The variable time step scheme controls the step size according to the general ideas proposed by \cite{Ces61,HNW09}, and adapted to our particular case
 with parameters taken from \cite{PFTV92}. The result of an integration at time $n+1$ is accepted, depending on the estimated error $E$ for the
 fields.  The error estimation  $E$ is based on the difference between the solution obtained with a third  and a second order scheme. Then essentially
 the new time step is evaluated as follows:
 $$h_{new}= s\left(\frac{E}{\text{tolerance}}\right)^{-1/(q+1)}h_{old}.$$ In practice, this expression is tuned and a maximum increase of the step size is allowed.
 Here $s$ is a safety factor, and $q$ is the order of the numerical scheme.
Acceptance of the result of an integration means that $E$ is below a certain  tolerance, which  is  explained in the subsection \ref{S3.2}. If the result is accepted, a new time step
 is proposed according to the law:
 $$ h_{new}= \left\{ \begin{array}{ll}s h_{old} \left( \frac{E}{\text{tolerance}}\right)^p, & E>\text{tolerance}\cdot\left(\frac{5}{s}\right)^{1/p}\\
 5h_{old},& E\leq \text{tolerance}\cdot\left(\frac{5}{s}\right)^{1/p}, \end{array} \right.$$ where  $h$ is the size of the time step and $s=0.9,$ $p=-0.33$ $(q=2)$.
 In case of rejection, the step  is decreased as follows: $$h_{new}= s\left(\frac{E}{\text{tolerance}}\right)^{-1/(q+1)}h_{old},$$
with $q=3.$
\subsection{The fully implicit method}\label{S3.1}
BDFs are a particular case of multistep formulas which are  {\it implicit},  thus  the BDF scheme   implies solving at each time step the problem (see \cite{HNW09}):
$$
\partial_t {\bf y}^{n+1}={\bf f}({\bf y}^{n+1}),
$$
which in the  particular problem under consideration becomes:
 \begin{align}
 &0=\nabla\cdot \mathbf{u}^{n+1} \label{eqtem1}\\
 &0=R\theta^{n+1}\vec{e_3}-\nabla P^{n+1}+\text{div} \left(\frac{\nu(\theta^{n+1})}{\nu_0}(\nabla \mathbf{u}^{n+1}+(\nabla \mathbf{u}^{n+1})^T)\right)\label{eqtem2}\\
 &\partial_t \theta^{n+1}=-\mathbf{u}^{n+1}\cdot \nabla \theta^{n+1}+\Delta \theta^{n+1},\label{eqtem3}
 \end{align}
where  $\partial_t \theta^{n+1}$ is replaced by the expression (\ref{eqdttheta}). The solution to the system (\ref{eqtem1})-(\ref{eqtem3}) is  our  benchmark  for transitory and time-dependent regimes.

The nonlinear terms on the right-hand side of these equations   are approached  at each time step, $n+1$, by  a Newton-Raphson method similar to the one described before to find numerical stationary solutions. We
assume that the solution at time $n+1$ is a small perturbation $\tilde{Y}$ of an approximate solution.
Linear equations for $\tilde{Y}$ are derived by introducing the analogue of expression (\ref{pertu}) into the nonlinear  terms
  of equations  (\ref{eqtem1})-(\ref{eqtem3}) and cancelling all the nonlinear terms in tilda.
  The resulting linearized terms are the same as those
  appearing in Eqs.  (\ref{eqtem13})-(\ref{eqtem43}).  For the first step $s=0,$ we take as an approximate solution  the solution at time $n$. The unknown perturbation fields $\tilde{Y}$ are expanded by Eq. (\ref{eqexpansion3}), which is replaced in the equations  and boundary conditions at the collocation points according to the rules described in the previous section.  In particular, the improved boundary conditions for pressure, previously described, are used.  A linear system is obtained:
 \begin{equation}
Ay=b\label{linearsys}
 \end{equation}
 which is iteratively solved at each time step $n+1$ until the perturbation $\tilde{Y}$ is below  a tolerance.
 Here, $A$ is a matrix of order $4\times L\times M$ and $y$ is the vector containing the unknown coefficients of the fields $\tilde{Y}$.
 As previously performed the matrix $A$ is converted into a full rank matrix by removing  the projection of equation \eqref{eqtem2} at node $j=1$, $i=2$,  and fixing the pressure by adding at this point the equation $b_{10}^P=0$. This provides accurate results and is computationally  cheaper than the pseudoinverse method.

\begin{figure}[h]
   \centering
   \subfigure[Time evolution of a transitory regime]{\includegraphics[width=8cm]{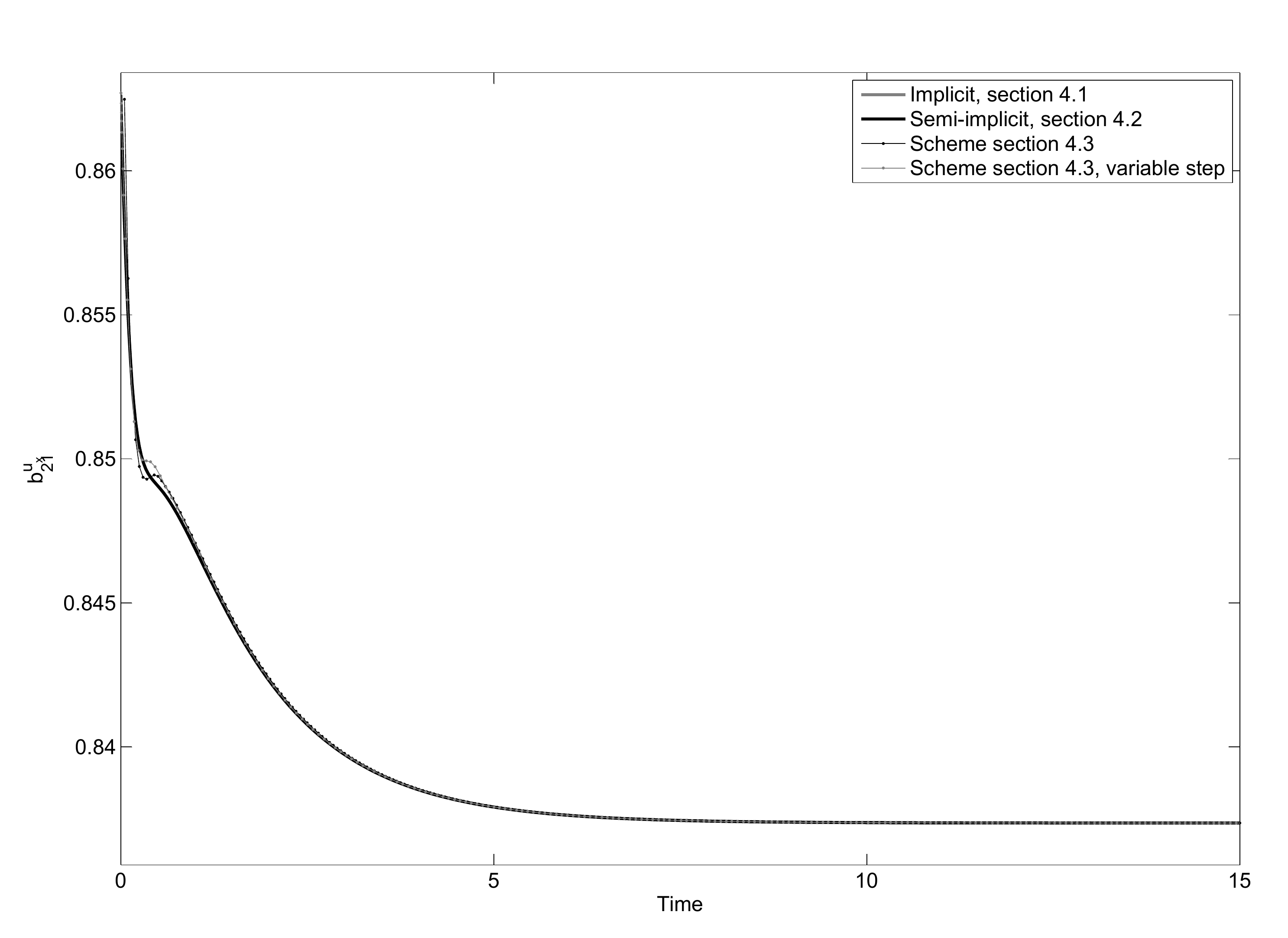} \label{FR75et}}
   \subfigure[Evolutions of the error versus time]{\includegraphics[width=8cm]{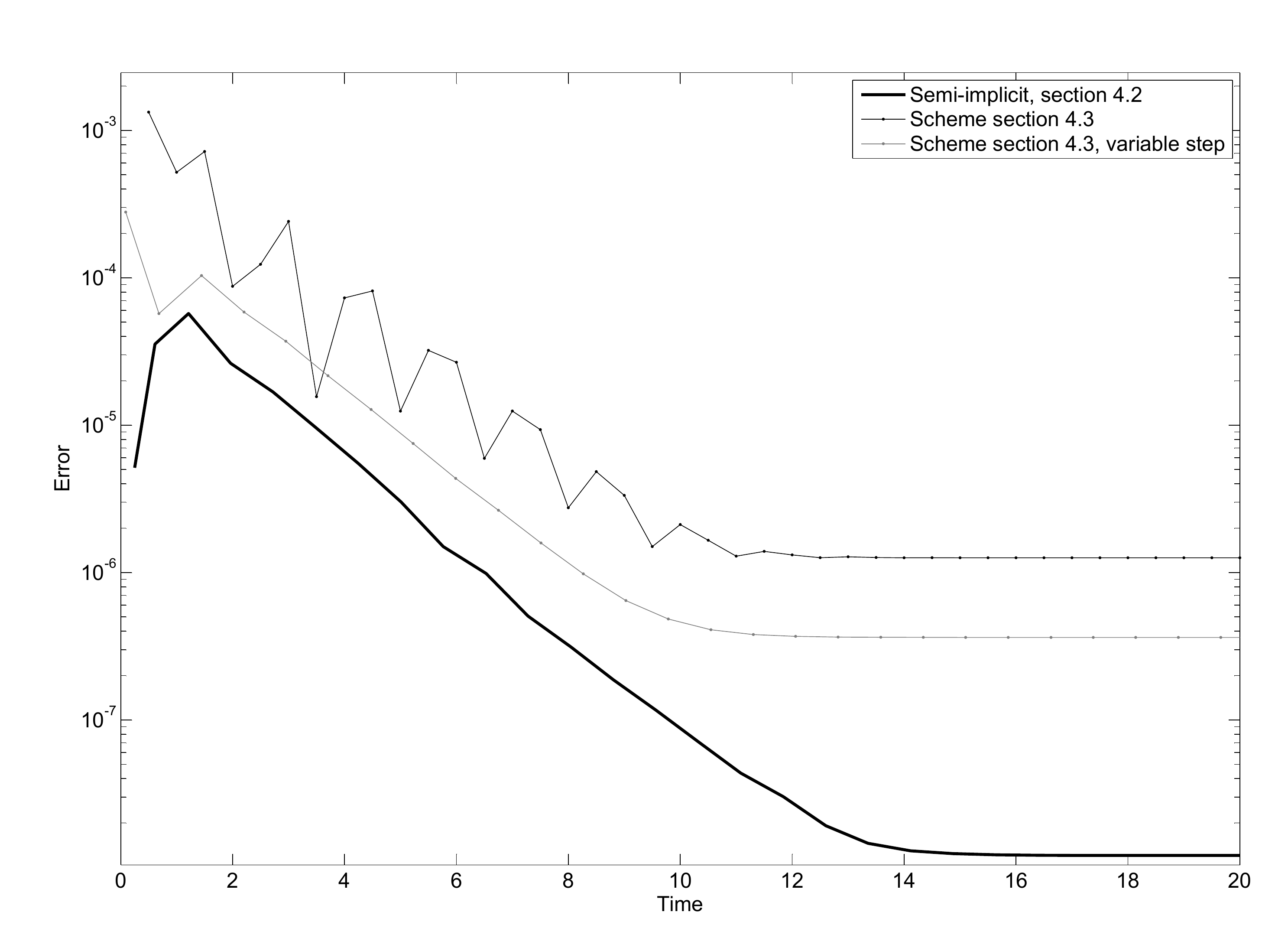}\label{FR75error}}
    \caption{Transition of an initial data to a stationary solution with $\Gamma=3.4$,
$R = 75$ for the exponential viscosity law \eqref{eqexp}.}\label{FR75}
   \end{figure}

 \subsection{The semi-implicit method}
 \label{S3.2}

Implicit methods are a robust and  numerically stable choice  for stiff problems. However,  they may be rather demanding computationally, since at each time step they require
several matrix inversions to solve the system (\ref{linearsys}) at successive iterations. This subsection proposes an alternative semi-implicit scheme that is computationally
 less demanding than the fully implicit scheme.

 Similarly to the fully implicit method, the semi-implicit scheme  approaches the nonlinear terms in Eqs.  (\ref{eqtem1})-(\ref{eqtem3}) by  assuming that the solution
at time $n+1$ is a small perturbation $\tilde{Z}$ of the solution at time $n$; thus,  ${\bf z}^{n+1}={\bf z}^{n}+\tilde{Z}$.
Once linear equations for $\tilde{Z}$ are derived, the equations are rewritten by replacing  $\tilde{Z}={\bf z}^{n+1}-{\bf z}^{n}$. Additionally, the linear system is completed by using expression   \eqref{eqdttheta}   for the time derivative of the temperature. The solution is obtained  at each step by solving the resulting linear equation for variables in time $n+1$. As before,
   the unknown fields are expanded by Eq. (\ref{eqexpansion3}), which is replaced in the equations  and boundary conditions at the collocation points according to the rules described in the previous section. At each time step the linear system:
 \begin{equation}
Ay=b
 \end{equation}
 is solved, where $A$ is a matrix of order $4\times L\times M$ and $y$ is the vector containing the unknown coefficients of the expansions of the ${\bf z}^{n+1}$ fields.
 The matrix is transformed into a full rank matrix with the same procedure used for the fully implicit case.

The fully implicit and the semi-implicit methods are implemented with  a variable time step scheme that  requires an error estimation
 based on the difference between the solution obtained with a third  and a second order scheme.  The second order scheme
 approaches the time derivative of the temperature  as follows:
 \begin{align}\label{eqdttheta2}
 \partial_t\theta^{n+1}=\frac{(\Delta t_n^2+2\Delta t_n\Delta t_{n+1})\theta^{n+1}-(\Delta t_n + \Delta t_{n+1})^2\theta^n + \Delta t_{n+1}^2 \theta^{n-1}}{\Delta t_n\Delta t_{n+1}(\Delta t_{n+1}+\Delta t_n)},
 \end{align}
 where $\Delta t_{n}=t_{n}-t_{n-1}.$ In practice, this changes the linear system to be solved at each step, as follows:
 $$\widetilde{A}\widetilde{y}=\widetilde{b}$$
 The computation of the second order solution thus leads to an additional  matrix inversion at each time step, and as a consequence the full calculation is slowed down considerably.
 To avoid this additional  inversion,  we estimate the error  by  measuring  instead how well the third order solution $y$ satisfies the second order system, i.e.:
 \begin{equation}\label{eqE} E=\frac{\| \widetilde{b}-\widetilde{A}y\|}{\| b\|} \end{equation}
where $ \| \cdot  \|$ represents the  $l^2$ norm. Acceptance of the result of an integration means that $E$ is below a tolerance that we fix at $5\cdot 10^{-6}$. Once the error is estimated,
the step size is determined as explained at the beginning of Section 4.

   \begin{table*}
  \centering
  \footnotesize
  \begin{tabular}{|c|cccc|}
    \hline
     & M=36 & M=38 & M=40 & M=42 \\
    \hline
    L=29 & 0.0030 - 0.0002i & 0.0046 & 0.0046 & 0.0047  \\
         & -8.2205 & -8.4419  & -8.4419   & -8.4419   \\
    \hline
    L=31 & 0.0017  & -0.0019  & 0.0017 & 0.0017  \\
         &  -8.4418  & -8.4417   &  -8.4418   & -8.4418  \\
    \hline
    L=33 & 0.0011  & 0.0011  & -5.79e-4 -1.11e-4i & 7.95e-4 + 6.30e-5i\\
         & -8.4418   & -8.4418   & -8.4417   &  -8.4418  \\
    \hline
  \end{tabular}
  \caption{Computation of the two eigenvalues with largest real part  for the stationary solution obtained  at $\Gamma=3.4$, $R=78$ at different expansions $L\times M$.}\label{t2}
\end{table*}

 \begin{table*}
  \centering
  \footnotesize
  \begin{tabular}{|c|cccccc|}
    \hline
     & M=30 & M=40 & M=50 & M=60 & M=70& M=80 \\
    \hline
    L=31 &  0.9433 + 0.7023i &   0.2227  &  0.5829 + 0.4345i & 0.3926            &  0.4217          & 0.2265 \\
         &  0.9433 - 0.7023i &  -2.9525  &  0.5829 - 0.4345i & -2.3225 + 0.7449i & -2.1744 + 0.7450i& -2.6555 + 0.7536i\\
    \hline
    L=37 & 0.3564          &   0.1837        &  0.1369      &  0.1318         & 0.1834&   0.0946\\
         & -2.1572 + 0.0829i &  -2.1273 + 0.0843i &  -2.6814     & -2.0368 + 0.0843i & -2.1427 + 0.0841i&  -2.1495+ 0.0964i\\
    \hline
    L=43 & -0.2039 + 0.0913i  &  0.1767              & 0.0773 +0.0371i & -0.0289 + 0.0886i & 0.0585& 0.0464\\
         & -0.2039 - 0.0913i  &  -2.7849 + 0.0546i   & 0.0773 -0.0371i & -0.0289 - 0.0886i & -2.1576+9.7059e-3i& -2.1308\\
    \hline
    L=49 & -0.1436           & 0.0836           & 0.0705 & 0.0379 &     9.7501e-3         & 4.7754e-3\\
         & -2.4633 +0.8885i &  -2.8147 + 0.0127i   & -2.3698 & -2.3625  &  -2.1391+ 6.3240e-4i & -2.1482\\
    \hline
    L=55 &   -0.1382           & 0.0571    &  0.0315               & 5.8407e-3            & 3.7262e-3& 7.8574e-4 \\
         &  -2.5619 + 0.8760i  &   -2.1569 &  -2.1893 + 1.0669e-3i & -2.1618 + 6.7549e-4i & -2.1597  & -2.1598\\
    \hline
    L=61 &   0.0913             &  0.0261 &  4.5781e-3    &  1.5021e-3   & 6.4589e-4 & 9.8951e-5\\
         &   -2.6323 + 0.1950i  & -2.1623 &  -2.1606      &   -2.1603    & -2.1599   & -2.1599 \\
    \hline
  \end{tabular}
  \caption{Computation of the two eigenvalues with largest real part  for the stationary solution obtained  at $\Gamma=3.4$, $R=110$ at different expansions $L\times M$.}\label{t3}
\end{table*}

 \begin{figure}[h]
   \centering
   \includegraphics[height=10cm, width=16cm]{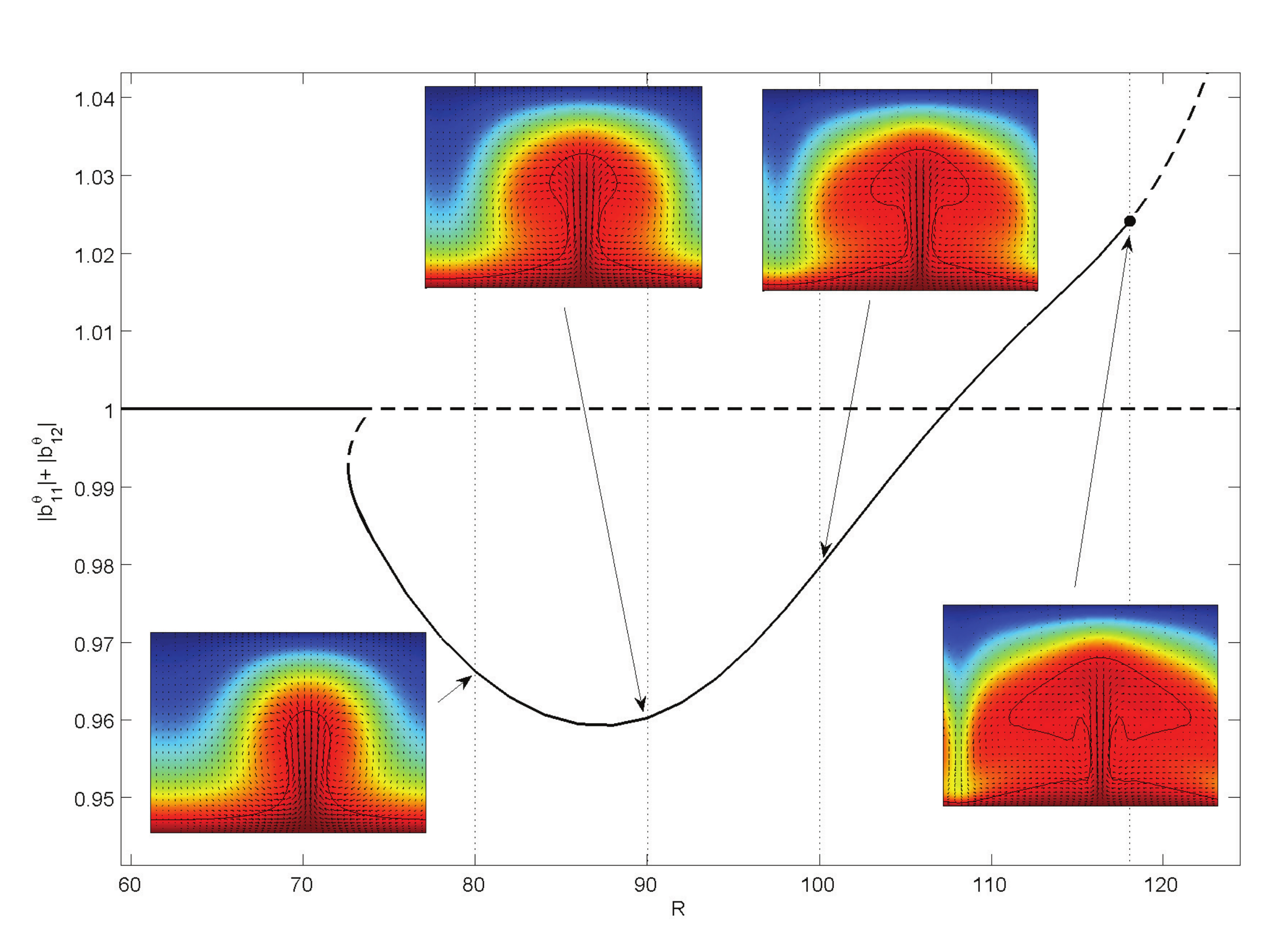}
    \caption{Bifurcation diagram for a fluid with viscosity dependent exponentially on temperature ($\mu=0.0862$ in law (\ref{eqexp})) at $\Gamma= 3.4$.
Stable branches are solid while unstable branches are dashed.
Solutions are displayed at the $R$ numbers highlighted with vertical dashed lines.}\label{Fbranchdiagram}
   \end{figure}
   
\subsection{Other semi-implicit schemes}\label{S3.3}

For completeness, we describe here alternative semi-implicit schemes that have been successfully used in fluid mechanics and convection problems with   constant viscosity {and finite Prandtl number}. Despite its success in many fluid mechanics set-ups,  this scheme is insufficient for our problem.

The adaptation of the numerical scheme described in \cite{HuRa98,MBA10} to the problem under study is as follows;  the semi-implicit scheme at each time step decouples the heat \eqref{eqproblem3} and the momentum equations \eqref{eqproblem2}. The time discretization of the heat equation is as follows:
\begin{equation}\label{eqtemporaltheta0}
\frac{3 \theta^{n+1}-4\theta^n+\theta^{n-1}}{2\Delta t}+ 2\mathbf{u}^n\cdot \nabla \theta^n- \mathbf{u}^{n-1}\cdot \nabla \theta^{n-1}= \Delta \theta^{n+1},
\end{equation}
where the time derivative of the temperature field has been evaluated with a  second order fixed step BDF formula.
Once $\theta^{n+1},$ is known the velocity and pressure at time $t_{n+1},$ are obtained by solving the following linear system in the unknown fields:
\begin{align}\label{eq26}
 &\nabla \cdot\mathbf{u}^{n+1}=0, \nonumber \\
 & \nabla P^{n+1} = R \theta^{n+1} \vec{e_3} + \text{div} \left( \frac{\nu(\theta^{n+1})}{\nu_0}\left( \nabla \mathbf{u}^{n+1} + (\nabla \mathbf{u}^{n+1})^T \right)\right).
\end{align}
We implement this scheme by expanding the unknown fields in the equations \eqref{eqtemporaltheta0} and  \eqref{eq26} with Eq. (\ref{eqexpansion3}). They  are solved at successive times $t_n$. The discrete version of \eqref{eqtemporaltheta0}
 is a linear system with $L\times M$ unknowns, while the discrete version of  \eqref{eq26} has $3 \times L\times M$ unknowns. The decoupled nature of the procedure gives a certain speed  advantage to this method as compared with the previous one. However, the results presented in the next section confirm that this method is not robust when applied to the differential algebraic equations under study. Increasing the order of the method or using a variable step technique does not improve the output provided by this approach.

\begin{figure}[h!]
   \centering
   \subfigure[Time evolution of a transitory regime]{\includegraphics[width=8cm]{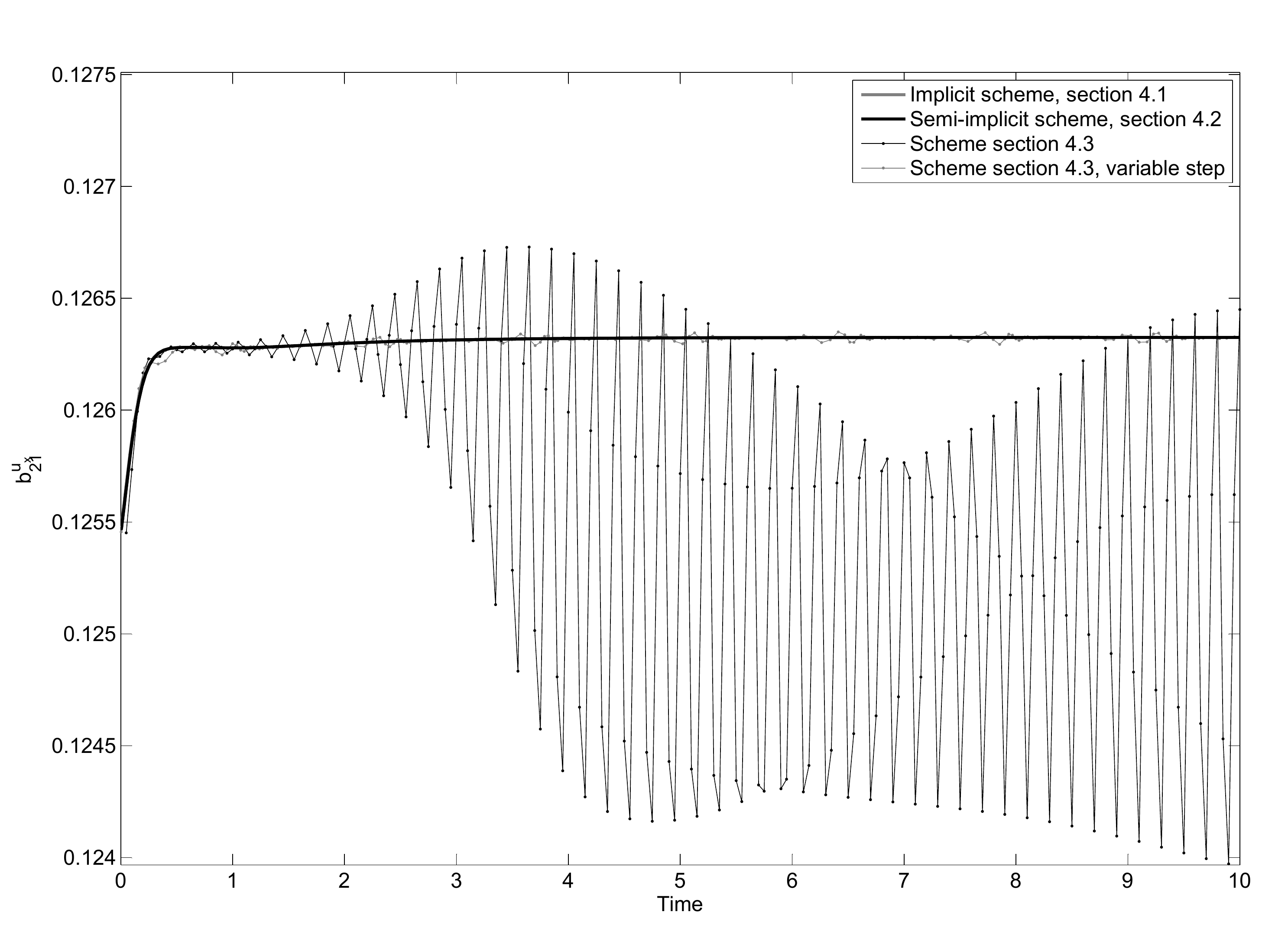} \label{FR78et}}
   \subfigure[Evolutions of the error versus time]{\includegraphics[width=8cm]{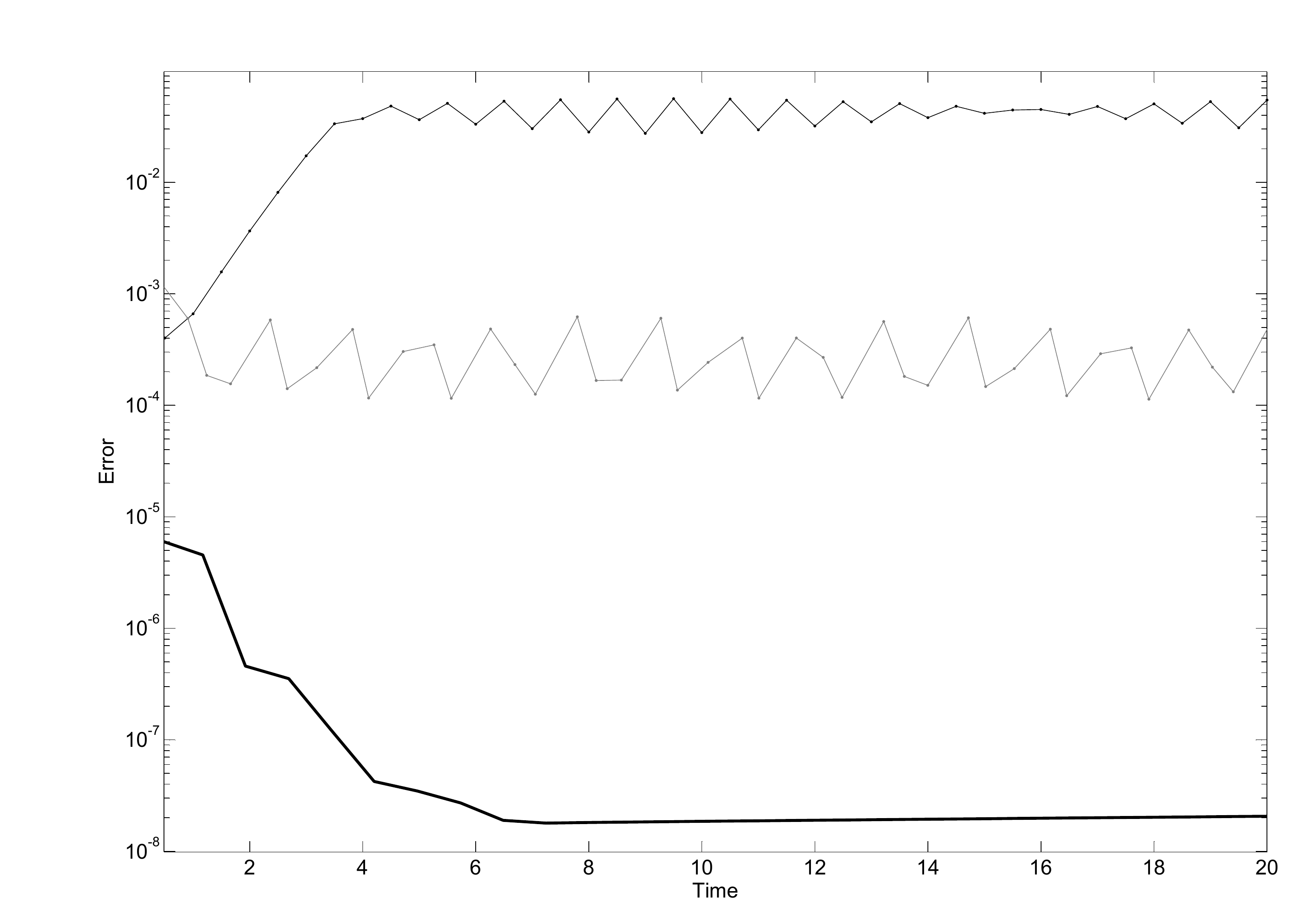}\label{FR78error}}
    \caption{Transition of an initial data to a stationary solution with $\Gamma=3.4$,
$R = 78$ for the exponential viscosity law \eqref{eqexp}. }\label{FR78}
   \end{figure}

   \begin{figure}[h!]
   \centering
   \subfigure[Time evolution]{\includegraphics[ height=8cm, width=8cm]{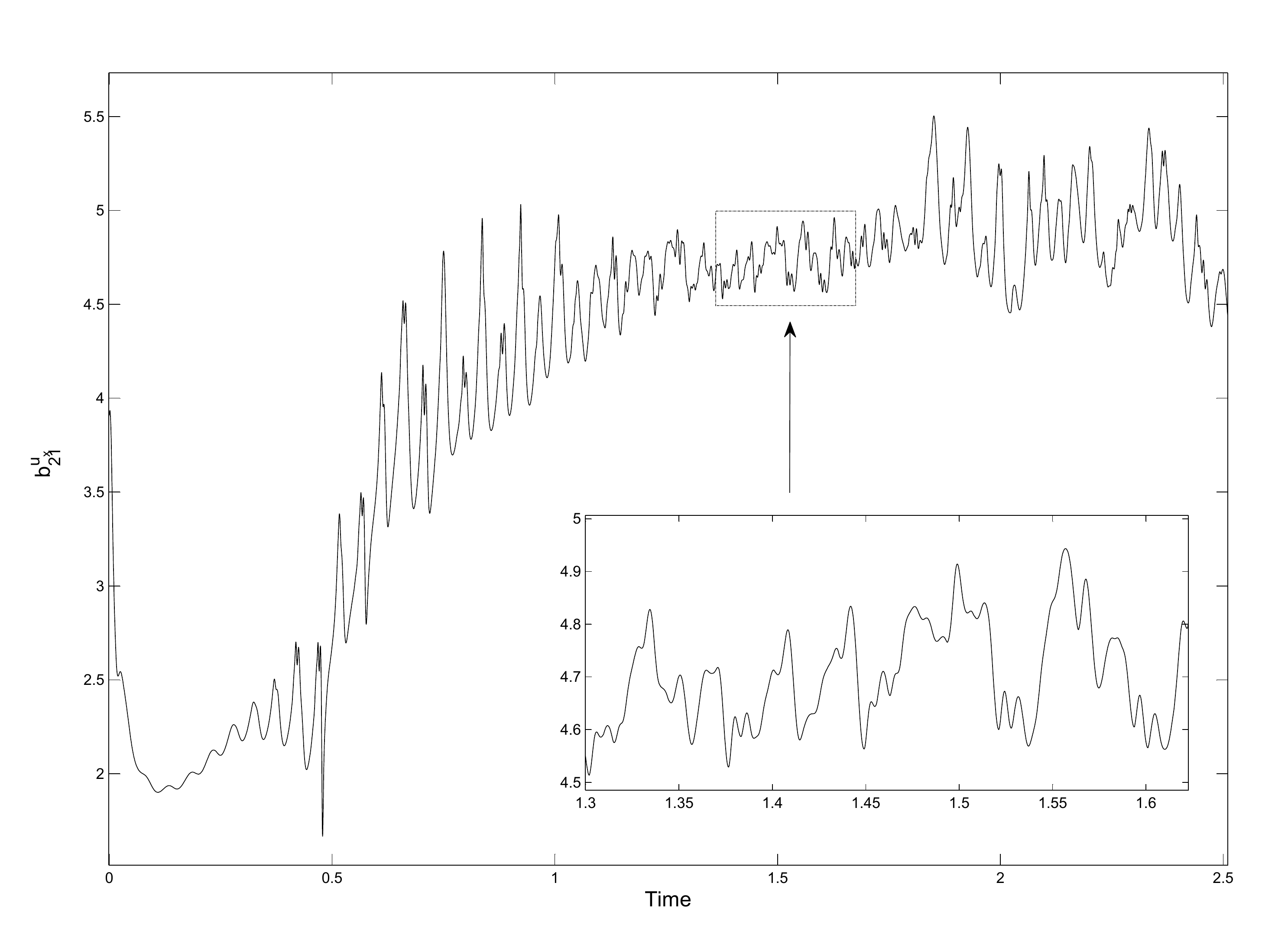}\label{FR120et}}
   \subfigure[Evolution of the error versus time]{\includegraphics[height=8cm, width=9cm]{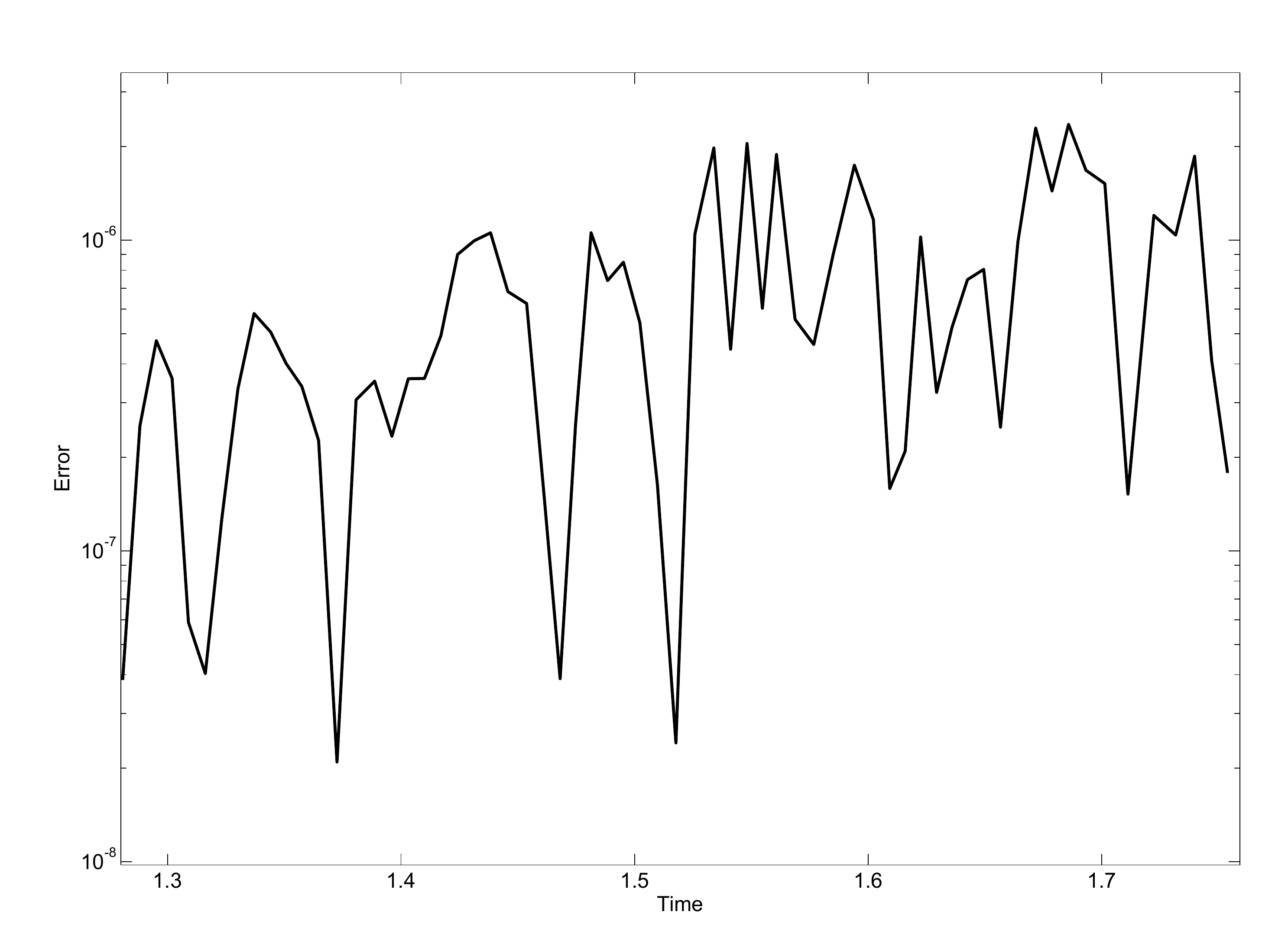}\label{FR120error}}
   \subfigure[Time $t=1$]{\includegraphics[height=5cm, width=5cm]{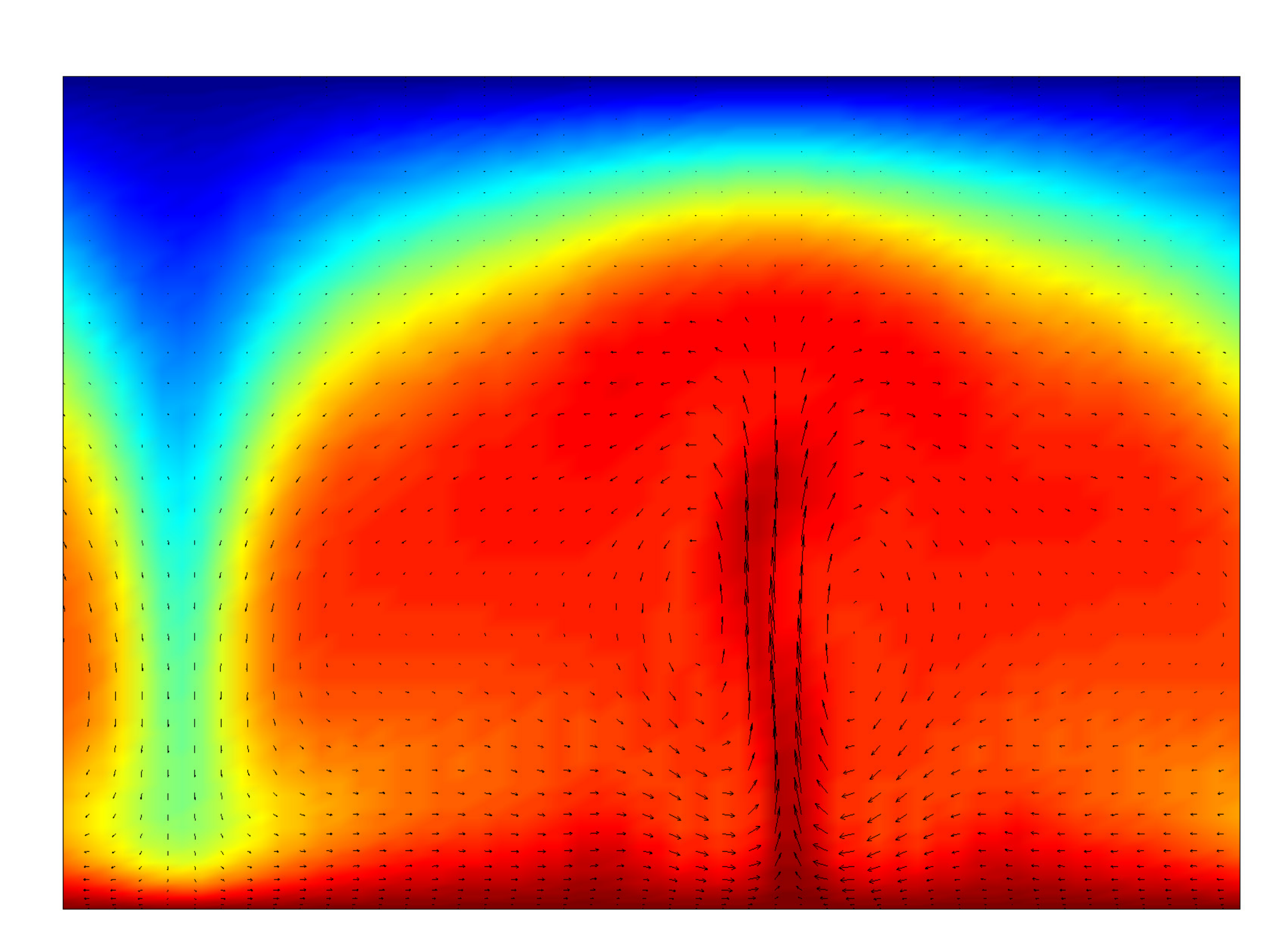}\label{FR120t1}}
   \subfigure[Time $t=1.1$]{\includegraphics[height=5cm, width=5cm]{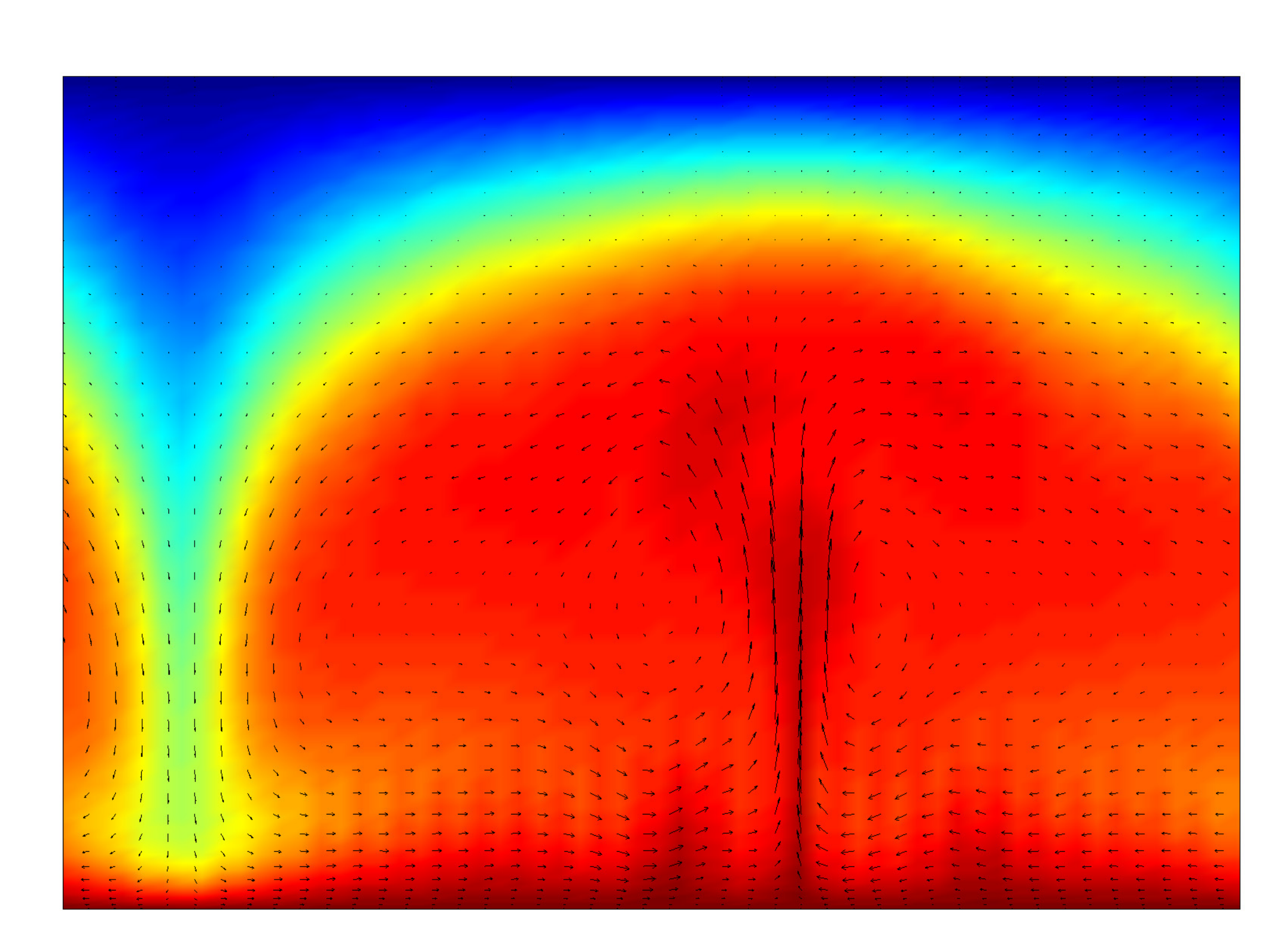}}
   \subfigure[Time $t=1.2$]{\includegraphics[height=5cm, width=5cm]{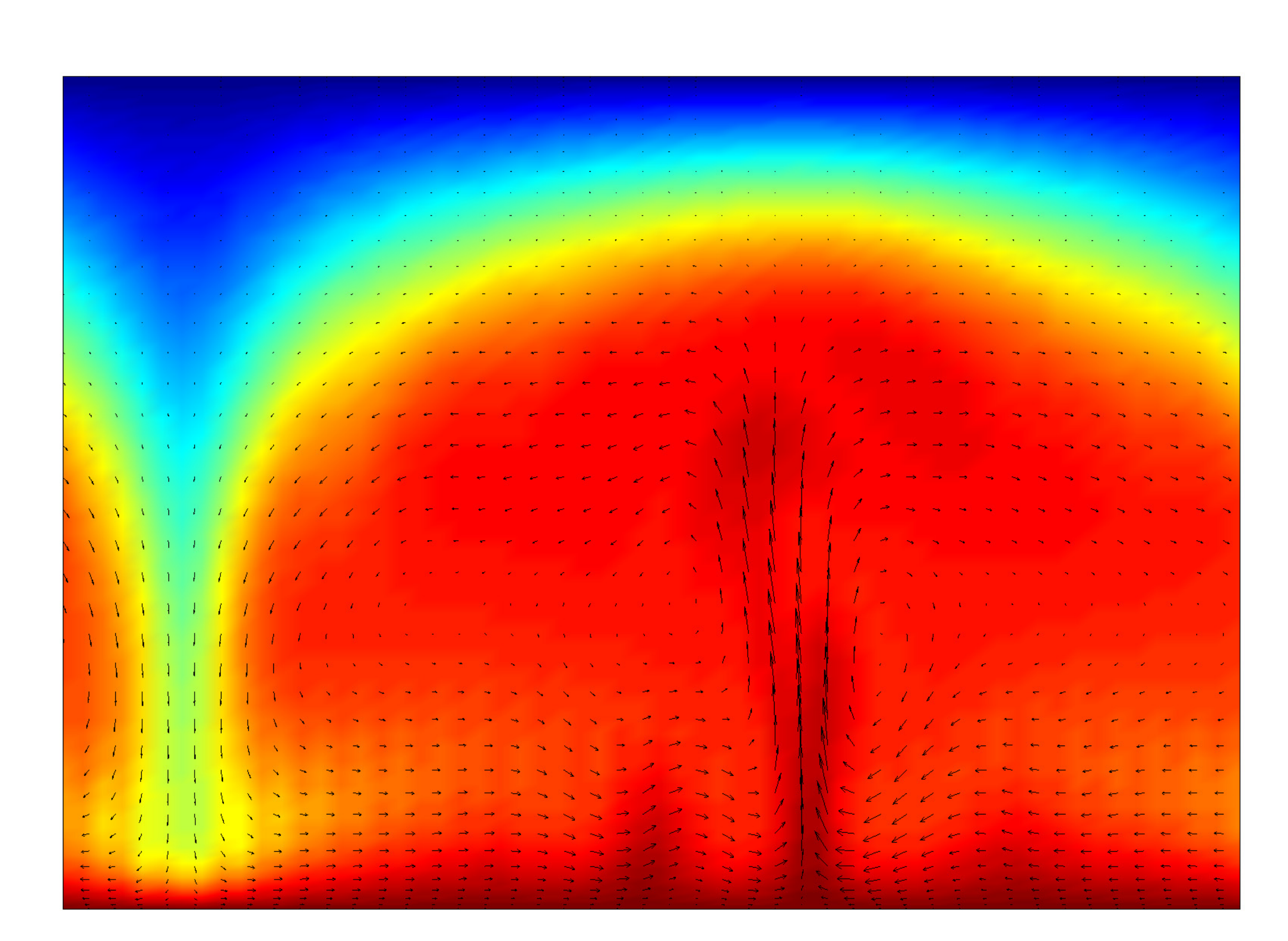}}
   \subfigure[Time $t=1.3$]{\includegraphics[height=5cm, width=5cm]{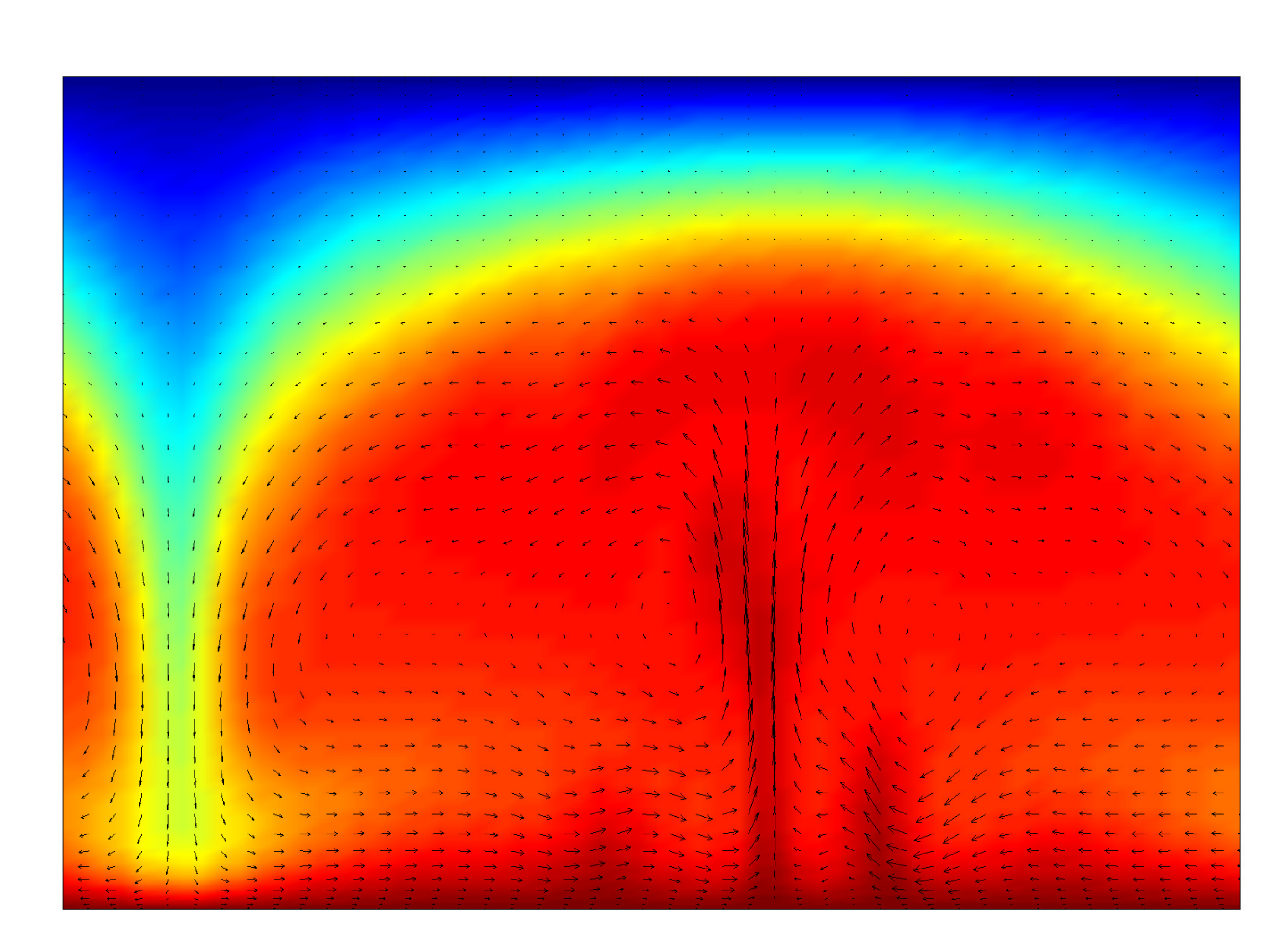}}
   \subfigure[Time $t=1.4$]{\includegraphics[height=5cm, width=5cm]{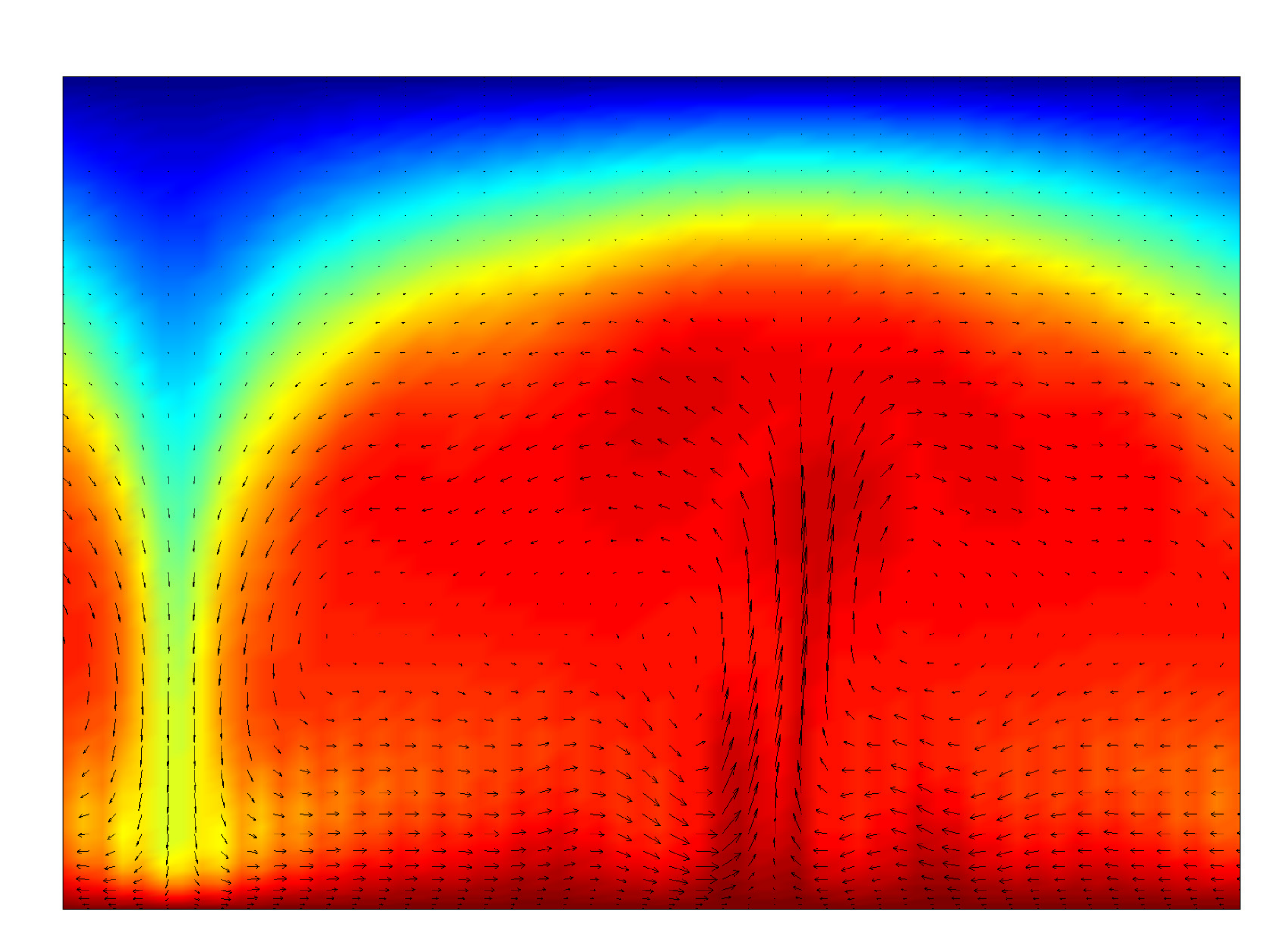}}
   \subfigure[Time $t=1.5$]{\includegraphics[height=5cm, width=5cm]{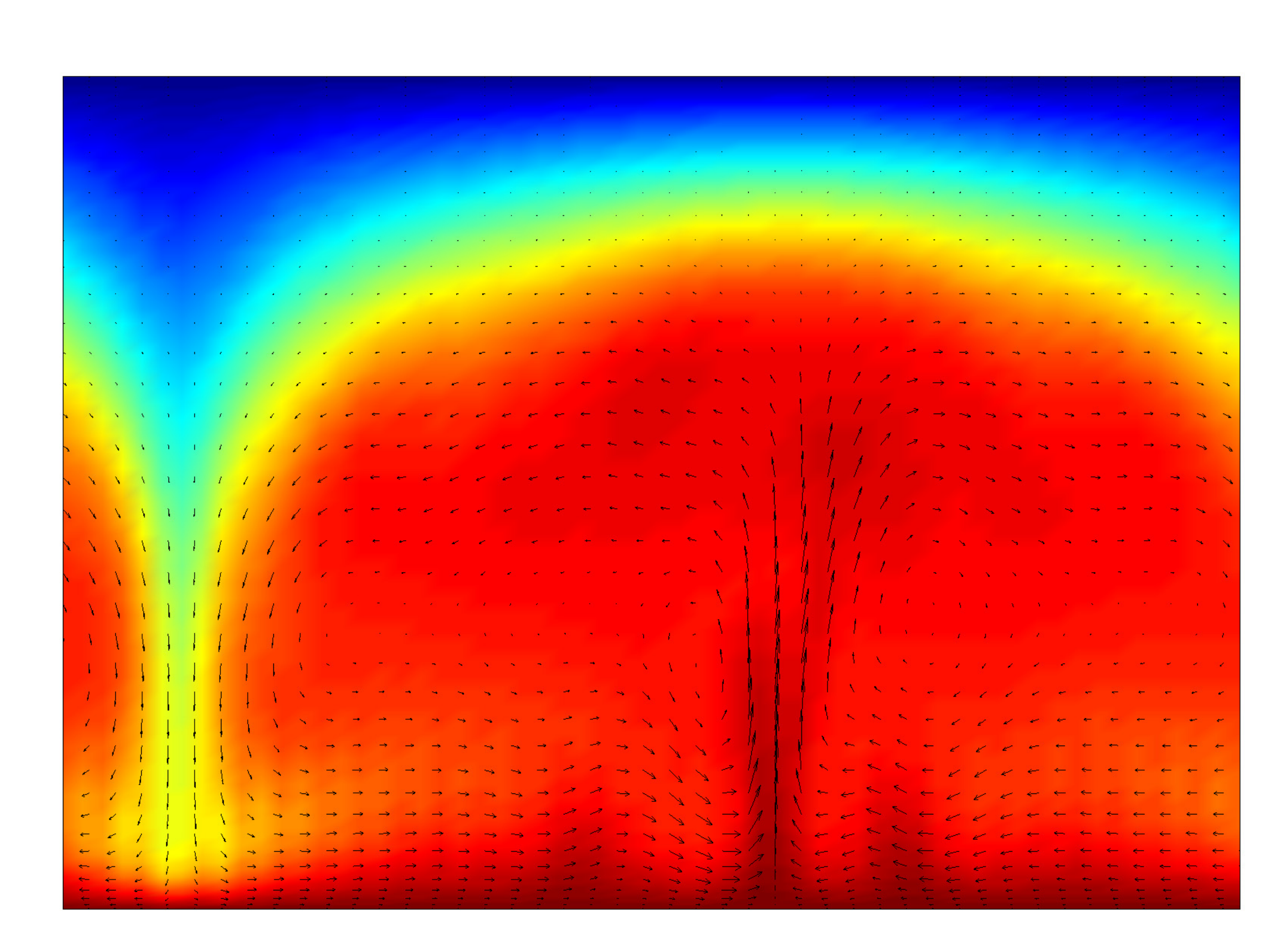}\label{FR120t1.5}}
    \caption{Temporal evolution of time dependent solution at $\Gamma=3.4$ and $R=120$ for the exponential viscosity law \eqref{eqexp}.}\label{FR120}
   \end{figure}

 \section{Results}
 \label{S4}

This section completes the description provided in \cite{PMH09} on the solutions of the convection problem in which the viscosity  depends exponentially on temperature.
 By describing  stationary, transitory and time dependent regimes for the  problem  under study,  the  consistency between the reported numerical  procedures is confirmed.

\subsection{Stationary solutions and their stability}

Figure \ref{Fbranchdiagram} displays the bifurcation diagram obtained  from the analysis of a fluid layer in a finite domain with aspect ratio  $\Gamma= 3.4$. Viscosity depends on the temperature  according to the exponential law  (\ref{eqexp}), with $\mu=0.0862$.  The viscosity contrast across the fluid layer depends on $R$ in such a way that at the instability threshold this contrast is around $6 \cdot 10^{2},$ while at the maximum Rayleigh numbers displayed in Figure  \ref{Fbranchdiagram}  it is of the order of $3.1 \cdot 10^{4}$. Stable branches are displayed as solid lines, while unstable branches are dashed.

The diagram displayed in  Fig. \ref{Fbranchdiagram}  is obtained by branch continuation techniques,
as explained  in \cite{PMH09}. The solutions are obtained with the procedure described in Section 3.2.
The vertical axis represents the sum $|b_{11}^\theta|+|b_{12}^\theta|$. These are coefficients  obtained from the expansion of the temperature field:
   \begin{align*} \theta (x,z,t)=& \sum_{l=1}^{\lceil L/2\rceil}\sum_{m=0}^{M-1} b^\theta_{lm}(t)T_m(z)\cos((l-1)x) \\ &+ \sum_{l=2}^{\lceil L/2\rceil}\sum_{m=0}^{M-1} c^\theta_{lm}(t)T_m(z)\sin((l-1)x).\end{align*}
Most of the coefficients $b_{lm}^{\theta}(t), c_{lm}^{\theta}(t)$ in this expansion are approximately zero, but others are not.  As a representation of the whole spatial function, we select the sum of the significant coefficients  $|b_{11}^\theta|$ and $|b_{12}^\theta|$.

Table  \ref{t2} confirms the convergence of the eigenvalues for a stationary solution obtained at  $\Gamma=3.4$, $R=78$. At this low $R$ number, results have two
significant decimal digits for expansions $L=29 \times M=38 $ onwards. The results  confirm that this is a  stable solution.
The   presence of a  zero eigenvalue   is expected due  to the symmetry SO(2) derived from the periodic boundary conditions.  Periodic boundary conditions imply that arbitrary  translations of the solutions  along the $x$-coordinate must also be solutions to the system. Thus,      instead of an isolated fixed point  at the bifurcation threshold a circle of fixed points emerges\cite{DM97}. The neutral direction is the direction  connecting  fixed points on this circle.
 Table \ref{t3} shows the convergence results obtained for $R=110$. At this larger $R$ number  higher expansions are required; this is also expected  because at large $R$ numbers the viscosity contrast across the fluid layer increases.  For expansions $L=55\times M=60$ onwards, a significant number of decimal digits is obtained.

Results displayed in  Figure \ref{Fbranchdiagram} are interpreted as follows: at the bifurcation threshold the fluid undergoes a subcritical bifurcation, as reported in \cite{W88,SOB82,PMH09,S12}. {The instability threshold of the unstable branch from the conductive solution occurs at $R=73.7544$ which coincides with the prediction of the linear theory $R=73.7501$ within a $0.006\%$. These  results confirm the high  accuracy of spectral methods  which is above for instance the outcome  reported in \cite{BMM92}   where using   finite volume methods  obtain on similar thresholds an accuracy of 0.4$\%.$ }
The unstable branch bifurcates at $R\sim74$, below the critical threshold for the conductive solution,  in a saddle-node bifurcation at which a stable stationary branch emerges. The pattern of the plume at this aspect ratio, consistent with diagram  \ref{Fexp} has wave number $m=1$. The stable branch becomes unstable at $R\sim118$ through a Hopf bifurcation.
Stationary solutions are displayed at the different $R$ numbers highlighted with vertical dashed lines. These results extend those reported in \cite{PMH09}, where the morphology of the plume has not been discussed.   Images in Fig.  \ref{Fbranchdiagram} show the evolution of the plume with the Rayleigh number.
As reported in \cite{KK97},  three idealized shapes for plumes are typically found: spout-shaped, where the tail of the plume is nearly as large
as the head; mushroom-shaped; and balloon-shaped. Figure \ref{Fbranchdiagram} confirms that at low $R$ numbers ($R=80$) the plume is spout-shaped.   At higher  $R=90,100$
the plume is more rounded at the top and becomes closer to a balloon-shaped plume while at $R=118$ the plume becomes closer to a mushroom-shaped one. Regarding the velocity field,
Moresi and Solomatov report in \cite{MS95} that from $10^4-10^5$ viscosity contrasts a stagnant lid develops, and   the upper part of the fluid, where the viscosity is much larger, does not move. The velocity fields overlapping the temperature patterns in Figure   \ref{Fbranchdiagram} confirm that at the larger viscosity contrasts obtained at $R=118,$ the  velocity in the upper part of the fluid is almost null.

The set of stationary solutions obtained with the techniques reported in Section 3, as already  noted by Pl\'a et al. in \cite{PMH09}
 is more comprehensive than what one would expect from mere direct time evolution simulations, because the latter  do not prove anything about  the asymptotic regime.  However, as also noted by these authors,  simulations of the evolution in  time  are also necessary to describe time-dependent regimes which are present at high $R$ numbers. From the computational point of view, the Newton-Raphson method is more advantageous than the time evolution schemes, as it finds a stationary solution in less than 50 iterations, while
 the semi-implicit scheme needs around 200 iterations (matrix inversions) to find the same  solution beginning from the same initial data.


\subsection{Time dependent and transitory regimes}

Figure \ref{FR75et} represents  the time evolution of the coefficient $b^{u_x}_{21}$ in  a transitory regime towards a fixed point.  A rescaled Time$=10t$ is represented on the horizontal axis, where $t$ is the dimensionless time. The simulation is produced  at $R=75,$
 which is above the instability threshold of the conductive solution, as confirmed in Figures \ref{Fexp} and \ref{Fbranchdiagram}.
The results displayed in  Figure \ref{FR75et} are obtained with all the numerical schemes described in Section 4.
 Figure \ref{FR75error} represents the evolution of the error versus time for the different schemes. The error is defined with respect to the benchmark solution obtained with the fully implicit approach.

Despite the good performance of all schemes at low $R$ number,  at higher $R$ numbers  the semi-implicit scheme reported in Section 4.3 breaks.
Fig \ref{FR78et} confirms this point by depicting at $R=78$ the time evolution of the coefficient $b^{u_x}_{21}$  in  a transitory regime towards a fixed point.
The schemes in Section \ref{S3.3} fail to solve the transition with fixed and variable time steps, which is nevertheless well determined  with the implicit and semi-implicit scheme in Section \ref{S3.2}.
Fig \ref{FR78error} reports the evolution of the error with respect the benchmark solution for the different schemes.

Figure \ref{FR120} confirms the existence of time-dependent convection after the  Hopf bifurcation occurred at $R=118$.
Fig \ref{FR120et} displays the time  evolution of the coefficient $b_{2m}^{u_x}$ versus time. { The dynamics is rather chaotic as is foreseen for high Rayleigh numbers. This is truly the case as in our study as the viscosity $\nu_0$ used to define $R$   is the maximum viscosity  in the fluid layer.  A redefined $R$ number from the smallest viscosity, as used other studies, would be  four order of magnitudes bigger than this one. }
Snapshots of the plume in the time-dependent regime are shown in Figures \ref{FR120t1}--\ref{FR120t1.5}. In this time series, hot blobs
ascend in the central part of the plume and these are released in the upper part of the fluid.
Fig \ref{FR120error} shows the evolution of the error  for the semi-implicit scheme.

{The time dependent solutions displayed in Figure  \ref{FR120} although confirms the validity of the method  do not particularly exhibit the influence of the symmetry, and thus
 the proposed spectral scheme does not show its power on this respect for the chosen test problem.
  However for other viscosity dependencies as the ones reported  in \cite{CM13}
  the current spectral method has successfully  described solutions whose existence is related  to the presence of the  symmetry.  }

As regards  computational performance, the semi-implicit  scheme requires higher expansions  than the fully implicit scheme in order to achieve  stability, but as it eventually requires fewer matrix inversions --both because  it requires only one inversion at each time step and because larger step sizes are allowed-- it is slightly  faster than the fully implicit scheme.  {ÊOn average, for the simulations reported in this article the semi-implicit scheme requires 80 time units of time
for doing what   the fully implicit scheme takes 100 time units. Thus regarding CPU time the semi-implict method  is more advantageous.}

 \section{Conclusions}
 \label{S5}

This paper addresses the numerical simulation of time-dependent solutions of a convection problem with viscosity strongly dependent on temperature at infinite Prandtl number.
We propose a spectral method which deals with the primitive variables formulation. Time derivatives are evaluated by   backwards differentiation formulas (BDFs), which are adapted to perform with variable time step.  BDFs are a particular case of multistep formulas which are implicit. We solve the fully implicit problem and compare it with a semi-implicit method. For the problem under study, the proposed semi-implicit method is shown to be accurate and to have a slightly faster  performance than the fully implicit scheme.
We further show  that other semi-implicit schemes, which provide a good performance in classical convection problems with constant viscosity and finite $Pr$ number, do not
succeed in this set-up.

The time-dependent scheme succeeds in completing the results reported for this problem by \cite{PMH09}.  Assisted by bifurcation techniques,  we have gained insight into the possible stationary solutions satisfied by the basic equations. The morphology of the plume is described and compared with others obtained in the literature.
Stable stationary
solutions become unstable through a Hopf bifurcation, after which the time-dependent regime is solved by the spectral techniques proposed in this article.

 {The time dependent solutions found for the exponential viscosity  law   do not   evidence the influence of the symmetry.  However in \cite{CM13} for  different  viscosity dependences, at high-moderate viscosity contrasts, it is reported that the proposed scheme is   successful to this end. Finite element methods and finite volume methods have proven to be successful to reach  extremely large viscosity contrasts up to 10$^{10}$-10$^{20}$ and we have not improved these limits with spectral methods.  However at moderate viscosity contrasts the purpose of later techniques  seems to be justified by novel  dynamical evolutions derived from the presence of symmetries   where they can play a better role  than other discretizations and become the standard method as has been the case in classical convection problems with constant viscosity \cite{T89,MBA10,GNG-AS10,MBH97}.}

 \section*{Acknowledgements}

We thank A. Rucklidge and J. Palmer for useful comments and suggestions. We are grateful to CESGA and to CCC of Universidad Aut\'onoma de Madrid for computing facilities.
This research is supported by the Spanish Ministry of Science under grants  MTM2008-03754, MTM2011-26696 and MINECO: ICMAT Severo Ochoa project SEV-2011-0087.


\bibliographystyle{unsrt}
\bibliography{local}







\end{document}